\documentclass[11pt]{article}
\DeclareMathAlphabet{\scr}{U}{rsfs}{m}{n}

\pdfoutput=1
\usepackage{latexsym}
\usepackage{epsfig}
\usepackage[mathscr]{eucal}
\usepackage{amsfonts}
\usepackage{amscd}
\usepackage{cite}
\usepackage{array}
\usepackage{amssymb}
\usepackage{colordvi}
\usepackage[centertags]{amsmath}
\usepackage{enumerate}
\usepackage{graphicx}
\usepackage{booktabs}
\usepackage{theorem}
\usepackage[footnotesize]{caption}
\usepackage{soul}
\usepackage{mcite}
\usepackage{slashed}
\usepackage{xcolor}
\usepackage{bbm}
\usepackage[utf8]{inputenc}
\usepackage{fancyvrb}% provides improved Verbatim command
\usepackage[colorlinks,urlcolor=black,linkcolor = black,citecolor = black,]{hyperref}
\newcommand{\newc}{\newcommand}
\newc{\be}{\begin{equation}}
\newc{\ee}{\end{equation}}
\newc{\bea}{\begin{eqnarray}}
\newc{\eea}{\end{eqnarray}}
\newc{\ol}{\overline}
\newc{\wt}{\widetilde}
\newc{\bs}{\boldsymbol}
\newc{\m}{\mathcal}
\newc{\la}{\langle}
\newc{\ra}{\rangle}

\newcommand{\beq}{\begin{eqnarray}}
\newcommand{\eeq}{\end{eqnarray}}
\newcommand{\bpmatrix}{\begin{pmatrix}}
\newcommand{\epmatrix}{\end{pmatrix}}

\renewcommand{\ol}{\text{1l}}

\renewcommand{\eqref}[1]{Eq.~(\ref{#1})}

\newcommand{\bc}{\begin{center}}
\newcommand{\ec}{\end{center}}

\def\be{\begin{equation}}
\def\ee{\end{equation}}
\def\ba{\begin{eqnarray}}
\def\ea{\end{eqnarray}}
\def\ra{\rightarrow}

\def\m#1{m_{#1}}

\def\ltap{\;\centeron{\raise.35ex\hbox{$<$}}{\lower.65ex\hbox{$\sim$}}\;}
\def\gtap{\;\centeron{\raise.35ex\hbox{$>$}}{\lower.65ex\hbox{$\sim$}}\;}

\begin{document}

\title{
\vspace*{-3cm}
\phantom{h} \hfill\mbox{\small KA-TP-35-2017}%\\[-1.1cm]
%\vspace*{1cm}
\\[1cm]
%\vspace{13mm}
\textbf{High scale impact in alignment and decoupling in two-Higgs doublet models\\[4mm]}}

\date{}
\author{
Philipp Basler$^{1\,}$\footnote{E-mail:
\texttt{philipp.basler@kit.edu}} ,
Pedro M. Ferreira$^{2,3\,}$\footnote{E-mail:
  \texttt{pmmferreira@fc.ul.pt}} ,
Margarete M\"{u}hlleitner$^{1\,}$\footnote{E-mail:
\texttt{margarete.muehlleitner@kit.edu}} ,
Rui Santos$^{2,3,4\,}$\footnote{E-mail:
  \texttt{rasantos@fc.ul.pt}}\\[9mm]
{\small\it
$^1$Institute for Theoretical Physics, Karlsruhe Institute of Technology,} \\
{\small\it 76128 Karlsruhe, Germany}\\[3mm]
{\small\it
$^2$ISEL -
 Instituto Superior de Engenharia de Lisboa,} \\
{\small \it   Instituto Polit\'ecnico de Lisboa
 1959-007 Lisboa, Portugal} \\[3mm]
{\small\it
$^3$Centro de F\'{\i}sica Te\'{o}rica e Computacional,
    Faculdade de Ci\^{e}ncias,} \\
{\small \it    Universidade de Lisboa, Campo Grande, Edif\'{\i}cio C8
  1749-016 Lisboa, Portugal} \\[3mm]
{\small\it
$^4$LIP, Departamento de F\'{\i}sica, Universidade do Minho, 4710-057 Braga, Portugal}
}

\maketitle

\begin{abstract}
\noindent
The two-Higgs doublet model (2HDM) provides an excellent benchmark to
study physics beyond the Standard Model (SM).
In this work we discuss
how the behaviour of the model at high energy scales causes it to have a scalar with properties very similar
to those of the SM -- which means the 2HDM can be seen to naturally favor a decoupling or
alignment limit. For a type II 2HDM, we show that requiring the model
to be theoretically valid up to
a scale of 1 TeV, by studying the renormalization group equations
(RGE) of the parameters of the model,
causes a significant reduction in the allowed magnitude of the quartic couplings. This,
combined with $B$-physics bounds, forces the model to be naturally decoupled. As a consequence, any
non-decoupling limits in type II, like the wrong-sign scenario, are excluded.
On the contrary, even with the very constraining limits for the Higgs couplings from the LHC, the type I
model can deviate substantially from alignment. An RGE analysis similar to that made for type II shows, however,
that requiring a single scalar to be heavier than about 500 GeV would be sufficient for the model to be
decoupled.
Finally, we show that not only a 2HDM where the lightest of the CP-even scalars is the 125 GeV one does not require
new physics to be stable up to the Planck scale but this is also true when the heavy CP-even Higgs is the
125 GeV and the theory has no decoupling limit for the type I model.
\end{abstract}
\thispagestyle{empty}
\vfill
\newpage
\setcounter{page}{1}

%%%%%%%%%%%%%%%%%%%%%%%%%%%%%%%%%%%%%%%%%%%%%%%%%%%%%%%
\section{Introduction}
\label{sec:intro}
%%%%%%%%%%%%%%%%%%%%%%%%%%%%%%%%%%%%%%%%%%%%%%%%%%%%%%%

The discovery of the Higgs boson at the Large Hadron Collider (LHC)
by the ATLAS~\cite{Aad:2012tfa} and CMS~\cite{Chatrchyan:2012ufa} collaborations
has immediately triggered the discussion about which extensions of the
Standard Model (SM) could accommodate all data and still predict new physics,
observable during the Run 2 operation. One of the models that has been used
as benchmark for the searches for new physics by both ATLAS and CMS is the two-Higgs doublet model (2HDM)
in its CP-conserving, softly broken $\mathbb{Z}_2$ symmetric version. First proposed by T.D. Lee~\cite{Lee:1973iz},
2HDMs have been used as benchmark models not only for the LHC searches
but also theoretically. The different versions of the model allow for instance the introduction
of CP violation in the scalar sector, controlled flavour changing neutral currents or
dark matter candidates. The models also have a very different vacuum structure than the SM
one, because both charge and CP can be broken spontaneously. Furthermore, it is the
simplest extension where simultaneous minima of the same nature can occur.

With the mass of the Higgs boson determined with a good precision, the discussion
about the stability of the SM Higgs potential has restarted.
This involves studying the evolution of the SM quartic coupling
$\lambda$ with the renormalization group equations (RGE).
Two effects are here in play: on the one hand, the quartic coupling itself has a positive
contribution to its own RGE, and therefore tends to increase its value as one goes to higher energy scales;
on the other hand, the top quark Yukawa coupling has a negative
contribution to the RGE of $\lambda$, and tends
to reduce its value as one goes up in energy scale. As a result of these two effects, if the value of the
quartic coupling at the weak scale is too small, the RGE evolution will cause $\lambda$ to turn negative at
some point, and therefore the potential becomes unstable. If, however, the starting value of the quartic
coupling is too large, its RGE evolution will drive it to ever-higher
values so that the theory ceases eventually
to be perturbative and $\lambda$ develops a Landau pole. Prior to the Higgs boson discovery, these arguments
were used to constrain its mass~\cite{Hung:1979dn,Cabibbo:1979ay,Flores:1982rv,Lindner:1985uk,Sher:1988mj,
Lindner:1988ww,Ford:1992mv,Sher:1993mf,Isidori:2001bm,Einhorn:2007rv}. Now that we know its mass, we can verify
whether the potential
remains stable, and the theory perturbative, all the way up to the Planck scale. If that were not the case,
that would most likely be a sign of the existence of new physics, hitherto undiscovered, which would stabilize
the RGE evolution of the couplings.
It has been shown, in fact, that the SM vacuum is metastable if the theory is to be valid up to the Planck
scale~\cite{Bezrukov:2012sa, Degrassi:2012ry, Buttazzo:2013uya}. The only way  to have a stable electroweak
vacuum, according to these results, would therefore be for new physics to exist
at a scale well below the Planck scale.
The stability of the electroweak vacuum can be cured with the addition of
extra scalar degrees of freedom. With all the parameters of the SM determined,
the addition of a scalar singlet is enough to cure the
problem~\cite{Lebedev:2012zw, EliasMiro:2012ay, Pruna:2013bma, Costa:2014qga}.
As shown in~\cite{Costa:2014qga}, the addition of a complex singlet not only
provides a vacuum stable up to the Planck scale but in the broken phase of the
model one of the new scalars can have a mass below 125 GeV. For this particular model
only one scalar with a mass above 125 GeV is needed to stabilize the vacuum.
It has been shown, however, in the context of the SM, that the presence of new physics very close to the Planck scale can alter considerably such conditions of stability of the potential
~\cite{Branchina:2013jra,Branchina:2014usa,Branchina:2014rva},
and likewise eventual gravity contributions near the Planck scale can have
a sizeable impact ~\cite{Branchina:2016bws}.

The 2HDM belongs to the simplest extensions of the SM. An extra
scalar doublet enlarges the SM scalar model, but the remaining fields (gauge and
fermion) remain the same, as do the gauge symmetries of the model. A larger scalar sector implies
a more complex scalar potential -- indeed the version of the 2HDM potential in this work has 5 quartic couplings. And as in the SM,
one can ask whether the potential remains stable and perturbative, as one considers progressively
larger energy scales. As such, the RGE evolutions of the quartic
couplings of the 2HDM were studied by several
authors~\cite{Kreyerhoff:1989fa,Nie:1998yn,Kanemura:1999xf,Ferreira:2009jb} to ascertain the validity of the
model up to higher energy scales and, prior to the
Higgs discovery, to attempt to impose constraints on the unknown scalar masses of the model.
After the Higgs boson was discovered the stability of the several versions of the 2HDM was revisited in a number
of papers~\cite{Chakrabarty:2014aya, Dev:2014yca, Das:2015mwa, Chowdhury:2015yja, Ferreira:2015rha,
Chakrabarty:2016smc, Cacchio:2016qyh, Chakrabarty:2017qkh, Gori:2017qwg}. In all these works, the lightest
CP-even scalar is considered to be the discovered Higgs boson, and there is a common conclusion that, with all
relevant theoretical and experimental constraints taken into account, there always exists a region of the parameter
space where the 2HDM is valid up to the Planck scale.
Notice, however, that these studies assume a softly broken $\mathbb{Z}_2$ symmetry, the most popular
version of the 2HDM, and the region of parameter space found always included $m^2_{12} \neq 0$.
On the other hand, in reference~\cite{Chakrabarty:2014aya} a type II version
with an exact $\mathbb{Z}_2$ symmetric model was analysed, concluding that the potential  with $m^2_{12} = 0$
cannot be valid beyond 10 TeV without the intervention of new physics, a conclusion that was then confirmed
in later works. However, this conclusion is heavily dependent on the value of the charged
Higgs mass, $m_{H^\pm}$. We will show that the type II model with an exact symmetry is, when taking into
account the most recent bounds on $m_{H^\pm}$, in fact valid only up
to a few hundreds of GeV.

All studies quoted above agree on the fact that the quartic parameters of
the potential are increasingly small if the theory is to be valid up to higher and higher scales.
The issue of metastability in the 2HDM at high scales was discussed in~\cite{Chakrabarty:2016smc}
while in reference~\cite{Chakrabarty:2017qkh} it was shown that the heavy states of a 2HDM valid
up to Planck scale can be probed with a significance of at least 3$\sigma$ in the LHC high-luminosity run.
In this investigation we will not work in the exact alignment limit nor in the decoupling limit as
was done in previous works.
Our main goal will in effect be to show how 2HDM alignment may emerge ``naturally" from
requiring stability and perturbativity of the potential up to high energy scales.
We will furthermore scan over the entire parameter space
allowed by the most relevant up-to-date experimental
constraints. Moreover, we will take into account
the combination of all theoretical constraints, including the discriminant that forces
the minimum to be global\footnote{One-loop studies of the vacuum of
  some versions of the 2HDM were performed in~\cite{Ferreira:2015pfi, Cherchiglia:2017gko}.}, at various scales.
An interesting conclusion we will reach is that it is enough to have only
one heavy scalar boson to have decoupling (and therefore alignment) at
a scale as low as 1 TeV.
We will argue that decoupling can be defined for masses as low as 500 GeV -- in other words, if
even only one of the extra scalar masses is required to be above 500 GeV, the 2HDM with good
high-energy scale behaviour up to scales of the TeV order automatically has a scalar state with
SM-like properties.

We will analyse for the first time the stability of a softly broken $\mathbb{Z}_2$ symmetric 2HDM in the
case where the heaviest CP-even scalar is the 125 GeV Higgs boson -- the so-called heavy Higgs scenario. Quite
surprisingly, we will demonstrate that for this scenario there are regions of the parameter space for a
type I model for which the theory is well behaved
all the way up to the Planck scale. This is only possible if all the remaining scalar bosons have a mass
below 200 GeV. Obviously, due to the bound of 580 GeV on the charged Higgs mass for a type II model,
the heavy Higgs scenario in type II ceases to be valid already below an energy scale of about 1 TeV.

The paper is organized as follows. In section~\ref{sec:model} we describe the version
of the 2HDM used in this work and in section~\ref{sec:theoconstraints} we present
the theoretical constraints that we will impose at the various scales.
In section~\ref{sec:2hdmspace} we discuss the parameter space of the model in view
of the most relevant theoretical and up to date experimental constraints.
In sections \ref{sec:res1} to \ref{sec:res3} we present our results.
Our conclusions are given in Section~\ref{sec:concl}.
In appendix~\ref{RGE} we have collected the relevant RGEs for this study.

%%%%%%%%%%%%%%%%%%%%%%%%%%%%%%%%%%%%%%%%%%%%%%%%%%%%%%%
\section{The two-Higgs doublet model}
\label{sec:model}

The 2HDM is an extension of the SM in which the scalar potential is
built with two hypercharge 1 complex $SU(2)$ doublets $\Phi_1$ and $\Phi_2$.
When all the fields transform just as in the SM and no extra symmetries are imposed
on the Lagrangian, the most general Yukawa Lagrangian
gives rise to tree-level flavour-changing neutral currents (FCNC) which are
known to be severely constrained by experimental data.
Imposing a discrete symmetry on the scalar fields, $\Phi_1 \rightarrow \Phi_1$, $\Phi_2 \rightarrow - \Phi_2$,
and forcing the potential to be invariant under this $\mathbb{Z}_2$ symmetry, except for a dimension two
soft breaking term, the potential can be written as
\begin{eqnarray}
V(\Phi_1,\Phi_2) &=& m^2_1 \Phi^{\dagger}_1\Phi_1+m^2_2
\Phi^{\dagger}_2\Phi_2 - (m^2_{12} \Phi^{\dagger}_1\Phi_2+{\mathrm{h.c.}
}) +\frac{1}{2} \lambda_1 (\Phi^{\dagger}_1\Phi_1)^2 +\frac{1}{2}
\lambda_2 (\Phi^{\dagger}_2\Phi_2)^2\nonumber \\
&& + \lambda_3
(\Phi^{\dagger}_1\Phi_1)(\Phi^{\dagger}_2\Phi_2) + \lambda_4
(\Phi^{\dagger}_1\Phi_2)(\Phi^{\dagger}_2\Phi_1) + \frac{1}{2}
\lambda_5[(\Phi^{\dagger}_1\Phi_2)^2+{\mathrm{h.c.}}] ~. \label{higgspot}
\end{eqnarray}
We will work with a CP-conserving potential by considering all parameters of
the potential, together with the vacuum expectation values, to be real.
Also the parameter space we will consider is such that no spontaneous
CP breaking occurs.
This is in fact assured by simply requiring that a CP-preserving minimum exists~\cite{Ferreira:2004yd}.
When the symmetry is extended to the fermions in such a way that a
fermion of a given charge couples only to one doublet~\cite{Glashow:1976nt, Paschos:1976ay}
the Higgs
interaction terms become proportional to the quark mass terms and therefore
Higgs FCNC are absent at tree level. There are four independent choices
for the Yukawa Lagrangian~\cite{Barger:1989fj}. We will call type I
the model where only $\Phi_2$ couples to all fermions, type II the model
where $\Phi_2$ couples to up-type quarks and $\Phi_1$ couples to down-type quarks and leptons, Flipped (F) the model
where $\Phi_2$ couples
to up-type quarks and to leptons and $\Phi_1$ couples to down-type quarks
and finally Lepton Specific (LS) the model where $\Phi_2$ couples to all
quarks and $\Phi_1$ couples to leptons.

The two complex doublet fields $\Phi_1$ and $\Phi_2$ are expressed
in terms of charged complex fields $\phi_i^+$ and real  and imaginary
components of the neutral components of the doublets, $\rho_i$ and $\eta_i$ ($i=1,2$),
respectively. After electroweak symmetry breaking (EWSB) we can expand
the two doublets about their vacuum expectation
values (VEVs) $v_1$ and $v_2$ yielding
\beq
\Phi_1 = \left(
\begin{array}{c}
\phi_1^+ \\
\frac{\rho_1 + i \eta_1 + v_1}{\sqrt{2}}
\end{array}
\right) \qquad \mbox{and} \qquad
\Phi_2 = \left(
\begin{array}{c}
\phi_2^+ \\
\frac{\rho_2 + i \eta_2 + v_2}{\sqrt{2}}
\end{array}
\right) \;. \label{eq:vevexpansion}
\eeq
The mass matrices are the components of the bilinear terms
in the potential. As we assume charge and CP conservation we end up
with three $2 \times 2$ matrices ${\cal M}_S$,
${\cal M}_P$ and ${\cal M}_C$ for the neutral CP-even, neutral CP-odd
and charged Higgs sectors. The minimization conditions are
given by
\beq
\left.\frac{\partial V}{\partial \Phi_1}\right|_{\langle \Phi_i \rangle} =
\left.\frac{\partial V}{\partial \Phi_2}\right|_{\langle \Phi_i
  \rangle} = 0 \;, \label{eq:tadcond}
\eeq
which is equivalent to setting the two terms in the potential
linear in $\rho_1$ and $\rho_2$ to zero,
\beq
 m_{11}^2 - m_{12}^2
\frac{v_2}{v_1} + \frac{\lambda_1 v_1^2}{2}
+ \frac{\lambda_{345} v_2^2}{2} \label{eq:tad1} &=& 0 \nonumber \\
m_{22}^2 - m_{12}^2
\frac{v_1}{v_2} + \frac{\lambda_2 v_2^2}{2} + \frac{\lambda_{345}
  v_1^2}{2} \label{eq:tad2} &=&  0\;,
\eeq
where we have defined
\beq
\lambda_{345} \equiv \lambda_3 + \lambda_4 + \lambda_5 \;.
\eeq
These equations allow one to replace the $m_{11}^2$ and $m_{22}^2$ parameters by expressions in terms
 of the remaining parameters and the VEVs to obtain the following form for the mass matrices,
\beq
{\cal M}_S &=& \left( \begin{array}{cc} m_{12}^2 \frac{v_2}{v_1} +
    \lambda_1 v_1^2 & -
    m_{12}^2 + \lambda_{345} v_1 v_2 \\ -m_{12}^2 + \lambda_{345} v_1
    v_2 & m_{12}^2 \frac{v_1}{v_2} + \lambda_2
    v_2^2 \end{array}\right)
\label{eq:scalarmass} \\
{\cal M}_P &=& \left( \frac{m_{12}^2}{v_1 v_2} - \lambda_5 \right)
\left( \begin{array}{cc} v_2^2 & - v_1 v_2 \\ - v_1 v_2 &
    v_1^2 \end{array} \right)
\label{eq:pseudomass}
\\
{\cal M}_C &=& \left( \frac{m_{12}^2}{v_1 v_2} - \frac{\lambda_4 +
    \lambda_5}{2} \right) \left( \begin{array}{cc} v_2^2 & - v_1 v_2
    \\ -v_1 v_2 & v_1^2 \end{array} \right)  \;.
\label{eq:chargedmass}
\eeq
In~\eqref{eq:pseudomass} and \eqref{eq:chargedmass} we already see that
the pseudoscalar and charged scalar matrices have determinant equal to zero
--- and therefore a zero eigenvalue, corresponding to the expected
neutral and charged Goldstone bosons. The diagonalisation of the mass matrices
is performed via the following orthogonal transformations
\beq
\left( \begin{array}{c} \rho_1 \\ \rho_2 \end{array} \right) &=&
R(\alpha) \left( \begin{array}{c} H \\ h \end{array} \right)  \; , \label{eq:diagHh} \\
\left( \begin{array}{c} \eta_1 \\ \eta_2 \end{array} \right) &=&
R(\beta) \left( \begin{array}{c} G^0 \\ A \end{array} \right)  \;
, \label{eq:diagGA} \\
\left( \begin{array}{c} \phi_1^\pm \\ \phi^\pm_2 \end{array} \right) &=&
R(\beta) \left( \begin{array}{c} G^\pm \\ H^\pm \end{array}
\right) \label{eq:diagGHpm}
\;,
\eeq
where the rotation matrices have the form
\beq
R(\vartheta) = \left( \begin{array}{cc} \cos \vartheta & - \sin
    \vartheta \\ \sin \vartheta & \cos \vartheta \end{array} \right) \;,
\eeq
with $\vartheta = \alpha$ or $\beta$. These rotations lead us to the physical states,
which include one neutral CP-odd state, $A$, two neutral
CP-even states, $h$ and $H$, and two charged Higgs bosons,
$H^\pm$, besides the longitudinal components of the $W^\pm$ and the $Z$ bosons,
the pseudo-Nambu-Goldstone bosons $G^\pm$ and $G^0$, respectively.

The angle $\beta$ can be be defined at tree-level as
\beq
\tan \beta = \frac{v_2}{v_1} \;, \label{eq:tanbetadef}
\eeq
while $v_1^2 + v_2^2 = v^2 \approx (246 \mbox{ GeV})^2$
ensures the correct pattern of symmetry breaking. The mixing angle $\alpha$
can be written in terms of $({\cal M}_S)_{ij}$ ($i,j=1,2$), which are
the entries of the CP-even scalar mass matrix, as
\beq
\tan 2\alpha = \frac{2 ({\cal M}_S)_{12}}{({\cal M}_S)_{11}-({\cal M}_S)_{22}} \;.
\eeq
Introducing the quantity $M$ defined as
\beq
M^2 \equiv \frac{m_{12}^2}{s_\beta c_\beta} \;,
\eeq
with the short-hand notation $s_x \equiv \sin x$ etc., we can write
\cite{Kanemura:2004mg}
\beq
\tan 2\alpha = \frac{s_{2\beta} (M^2- \lambda_{345} v^2)}{c_\beta^2
  (M^2-\lambda_1 v^2) -s_\beta^2 (M^2-\lambda_2 v^2)}
\;.  \label{eq:alphadef}
\eeq
Finally, the scalar masses may be written as
\beq
m_{h,H}^2 &=& \frac{1}{2} \left[ ({\cal M}_S)_{11} + ({\cal M}_S)_{22} \mp
\sqrt{\left(({\cal M}_S)_{11} - ({\cal M}_S)_{22}\right)^2 + 4 (({\cal
    M}_S)_{12})^2} \right]  \nonumber \\
m_A^2 &=& M^2 -\lambda_5 v^2 \nonumber \\
m_{H^\pm}^2 &=& M^2 - \frac{\lambda_4+\lambda_5}{2} v^2 \;. \label{eq:masses}
\eeq

The potential has eight independent parameters and we choose: the four scalar masses
(the two masses of the CP-even states, the mass of the CP-odd state and the
mass of the charged Higgs boson), the rotation
angle in the CP-even sector, $\alpha$, the ratio of the vacuum expectation
values,  $\tan\beta=v_2/v_1$, the soft breaking parameter $m_{12}^2$ and
$v^2 = v_1^2 + v_2^2$. Without loss of generality, we adopt the conventions
$0\leq\beta\leq \pi/2$ and $- \pi/2 \leq  \alpha \leq \pi/2$.

The two doublets $\Phi_1$ and $\Phi_2$ are not physical fields, unlike the mass eigenstates.
This means that any linear combination of the doublets which preserves the form of the kinetic
terms of the theory is equally acceptable. This reparameterization
freedom implies that different
bases of the doublet fields can be chosen, without changing physical predictions of the model
and potentially simplifying the theory. It is sometimes useful to work in the so-called
Higgs basis, wherein one performs a $U(2)$ transformation on $\Phi_1$, $\Phi_2$ in such a manner
that only the first of the transformed fields, $\{ H_1, H_2 \}$, has a VEV. The Higgs basis may be
defined for our model by the rotation\footnote{The Higgs basis is in fact defined up to an arbitrary
complex phase multiplying the second doublet.}
\beq
\left( \begin{array}{c} H_1 \\ H_2 \end{array} \right) =
R_H \left( \begin{array}{c} \Phi_1 \\
    \Phi_2 \end{array} \right) \equiv
\left( \begin{array}{cc} c_\beta & s_\beta \\ - s_\beta &
    c_\beta \end{array} \right) \left( \begin{array}{c} \Phi_1 \\
    \Phi_2 \end{array} \right) \;,
\eeq
and hence the potential can be written as~\cite{Davidson:2005cw}
\begin{eqnarray}
V(H_1,H_2) & = & Y_1 H^{\dagger}_1 H_1 + Y_2 H^{\dagger}_2 H_2 -
(Y_3 H^{\dagger}_1 H_2+{\mathrm{h.c.}
}) +\frac{1}{2} Z_1 (H^{\dagger}_1 H_1)^2 +\frac{1}{2}
Z_2 (H^{\dagger}_2 H_2)^2 \nonumber \\
&& + Z_3 (H^{\dagger}_1 H_1)(H^{\dagger}_2 H_2) + Z_4 (H^{\dagger}_1 H_2)(H^{\dagger}_2 H_1) +
\{\frac{1}{2} Z_5(H^{\dagger}_1 H_2)^2 + [Z_6 H^{\dagger}_1 H_1 \nonumber \\
&& + Z_7 H^{\dagger}_2 H_2] H^{\dagger}_1 H_2
+{\mathrm{h.c.}}\} ~, \label{higgspotHB}
\end{eqnarray}
with the minimization conditions of the potential in this new basis implying that the
parameters $Y_3$ and $Z_6$ are related to one
another.

The reason why we are interested in this form of the potential is that it allows
to write expressions that facilitate in some cases the discussion of alignment and decoupling limits
in the 2HDM \cite{Gunion:2002zf}. Let us clarify what we mean by alignment and decoupling:
the LHC has shown beyond all doubts that the 125 GeV scalar which has been discovered has SM-like
behaviour -- meaning, it seems to couple to gauge bosons and fermions very much like the SM Higgs boson
would do. Within models with two doublets, this implies that the scalar state with 125 GeV mass
needs to be almost {\em aligned} with the VEV. How does one obtain such aligned regimes in the 2HDM?
The key issue is looking at the CP-even mass matrix
from \eqref{eq:scalarmass}: in the Higgs basis, this matrix becomes
\be
{\cal M}_S \,=\, \left( \begin{array}{cc} Z_1 \,v^2  & Z_6 \,v^2 \\ Z_6 \,v^2 & m^2_A + Z_5 \,v^2
\end{array}\right)\,.
\label{eq:mhhb}
\ee
Having an {\em aligned} scalar means that there won't be much mixing between the two CP-even states,
and this can be  achieved in two ways:
\begin{itemize}
\item One of the diagonal elements in \eqref{eq:mhhb} is much bigger than the other one. Since
$Z_1$ is a quartic coupling and therefore expected not to be large, this forces the (2,2) entry in
the matrix to be quite large, and it is simple to show that all extra scalars will be heavy.
In this regime, alignment is achieved in the {\em decoupling limit}.
\item The off-diagonal elements in \eqref{eq:mhhb} are much smaller than the diagonal
ones. In this regime, the masses of the extra scalars are not necessarily large, and the
SM-like behaviour of the 125 GeV state is said to be caused by the {\em alignment limit}.
\end{itemize}
Looking specifically at the couplings of $h$ or $H$ to gauge bosons, the relevant expressions
for our discussion are~\cite{Ferreira:2015rha,Gori:2017qwg}
\beq
|s_{\beta -\alpha} c_{\beta -\alpha}| =  \frac{|Z_6| v^2}{m_H^2 - m_h^2} \;, \label{eq:cba}
\eeq
and
\beq
Z_1 v^2 = m_h^2 s_{\beta -\alpha}^2 + m_H^2 c_{\beta -\alpha}^2\;, \label{eq:Z1mass}
\eeq
with $Z_1$ and $Z_6$ given in terms of the original parameters of the Lagrangian by~\cite{Davidson:2005cw}
\beq
Z_1 &=& \lambda_1 c_\beta^4 + \lambda_2 s_\beta^4 + \frac{1}{2} \lambda_{345}
s_{2\beta}^2\;, \label{eq:Z1} \nonumber \\
%\eeq
%
%\beq
Z_6 &=& - \frac{1}{2} s_{2 \beta} [\lambda_1 c_\beta^2 - \lambda_2 s_\beta^2 - \lambda_{345} c_{2\beta}] \;. \label{eq:Z6}
\eeq
Assuming that the lightest state is the one that is
aligned with the VEV, and that it has a mass of 125 GeV,
its tree-level couplings are very close to the SM Higgs ones.
This limit is attained by setting $c_{\beta -\alpha} \rightarrow 0$. Equation~(\ref{eq:cba}) tells us
then that it is sufficient to have $Z_6 \ll 1$ to be in the {\em alignment limit}.
In this regime, although the couplings of the 125 GeV Higgs are all
SM-like, the other Higgs bosons can in principle be light and
therefore be within the reach
of the LHC. To have alignment in the {\em decoupling limit}
the masses of the non-125 Higgs bosons must
not be much larger than 125 GeV. Defining a common mass scale
$m_{\phi_{\text{heavy}}}$ with $\phi_{\text{heavy}}=H, A$ and $H^\pm$ one can write~\cite{Kanemura:2004mg}
\beq
m_{\phi_{\text{heavy}}}^2 =  M^2 +
f(\lambda_i) \, v^2 + {\cal O} (v^4/M^2) \;, \label{eq:massrels}
\label{eq:kane}
\eeq
where $f(\lambda_i)$ denotes a linear combination of $\lambda_1...\lambda_5$.\footnote{In fact, we see from
Eqs.~(\ref{eq:masses}) that for $m_A$ and
$m_{H^\pm}$ the $v^4$ terms are not even present.} If $M^2 \gg f(\lambda_i) \, v^2$ all masses are of the order
of $M$ and therefore quite large --- and from eq.~\eqref{eq:Z6} again we obtain
$|\cos(\beta - \alpha)| \simeq 0$.

In the case $s_{\beta -\alpha} \rightarrow 0$ there is again alignment but now with the heavy CP-even
Higgs $H$, meaning this would correspond to the heavy Higgs scenario mentioned above.
The condition for this regime to occur is still $Z_6 \ll 1$, but now decoupling is not possible,
as the non-SM-like Higgs boson masses are not all much
larger than 125~GeV, in particular not $m_h$.

\section{Theoretical constraints on the parameters \label{sec:theoconstraints}}

The main goal of this study is to understand the effects of the RGE evolution
of the couplings,
from the weak scale (the mass of the $Z$ boson, $m_Z$) to higher scales all
the way up to the Planck scale, $\Lambda = 10^{19}$ GeV. Our procedure consists in
first inputting a set of 2HDM parameters at the weak scale
and verifying whether they satisfy the following theoretical demands:

\begin{itemize}
\item The potential is bounded from below, so that the theory is guaranteed to have a stable
vacuum of some sort.

This is achieved by demanding that the quartic couplings of the potential obey~\cite{Deshpande:1977rw, Klimenko:1984qx}
\begin{eqnarray}
\lambda_1 > 0 & , &  \lambda_2 > 0 \; ,\nonumber \\
\lambda_3 > -\sqrt{\lambda_1 \lambda_2} & , &
\lambda_3 + \lambda_4 - |\lambda_5| > -\sqrt{\lambda_1 \lambda_2} \;.
\label{eq:bfb}
\end{eqnarray}
These conditions have been shown to be necessary and sufficient~\cite{Ivanov:2006yq}
to ensure that the scalar potential is bounded from below (in the ``strong sense" as
defined in Refs.~\cite{Nagel:2004sw, Maniatis:2006fs}).
\item That the minimum is global and provides the right pattern of electroweak symmetry breaking.

Contrary to the SM, the 2HDM can have several simultaneous stationary points. Besides the CP-conserving
minimum, the model can have CP-violating (CPV) and Charge Breaking (CB) minima,
which are spontaneously generated. As shown in~\cite{Ferreira:2004yd, Barroso:2005sm,
Ivanov:2006yq,Ivanov:2007de},
if the potential is in a CP-conserving minimum, any other stationary point, if of a different nature
(either CPV or CB), is a saddle point with higher value of the potential.
Still, there is a possibility that two CP-conserving minima could co-exist. In this case tunnelling
could occur from our minimum to another one with a different
electroweak scale. In \cite{Barroso:2013awa, Barroso:2012mj} this
minimum was called the \textit{panic vacuum}. However, it was found
that verifying if the parameters of the potential
obey a simple condition~\cite{Barroso:2013awa,Barroso:2012mj, Ivanov:2008cxa} it is possible
to know exactly whether our CP-conserving vacuum is the global one. We define the discriminant
\be
D \,=\, m^2_{12} (m^2_{11} - k^2 m^2_{22}) \left(\frac{v_2}{v_1} - k\right)\;,
\label{eq:disc}
\ee
where $k = \sqrt[4]{\lambda_1/\lambda_2}$, and the VEVs are the ones that
define the correct pattern of symmetry breaking (meaning, they predict the correct gauge boson and fermion
masses, $v_1^2 + v_2^2 =$ (246 GeV)$^2$). The existence of a panic vacuum is thus summarised in the
following theorem:

{\em The vacuum with the correct pattern of symmetry breaking
is the global minimum of the potential if and only if $D > 0$.}
\item That perturbative unitarity\footnote{We note that a model that does not respect
perturbative unitarity is not necessarily wrong.
However, discussing this possibility is beyond the scope of this work.} holds.

We enforce tree-level perturbative unitarity by requiring that the
eigenvalues of the $2\to 2$ scalar scattering matrix are below
$8\pi$~\cite{Horejsi:2005da}. The full $2\to 2$ scattering
matrix of the fields in the gauge basis has been computed~\cite{Horejsi:2005da}
(see also~\cite{Kanemura:1993hm, Akeroyd:2000wc}), and its eigenvalues are
\beq
b_\pm &=& \frac{1}{2} \left( \lambda_1 + \lambda_2 \pm
 \sqrt{(\lambda_1-\lambda_2)^2 + 4 \lambda_5^2} \right)\nonumber \\
c_\pm &=& \frac{1}{2} \left( \lambda_1 + \lambda_2 \pm
 \sqrt{(\lambda_1-\lambda_2)^2 + 4 \lambda_4^2} \right)\nonumber \\
e_1 &=& \lambda_3 + 2 \lambda_4 - 3 \lambda_5 \nonumber \\
e_2 &=& \lambda_3 - \lambda_5 \nonumber \\
f_+ &=& \lambda_3 + 2\lambda_4 + 3 \lambda_5 \nonumber \\
f_- &=& \lambda_3 + \lambda_5 \nonumber \\
f_1 &=& \lambda_3 + \lambda_4 \nonumber \\
p_1 &=& \lambda_3 - \lambda_4 \;.
\eeq
The above eigenvalues are not all independent. As noted in
\cite{Horejsi:2005da},
\beq
3 f_1 &=& p_1 + e_1 + f_+ \\
3 e_2 &=& 2 p_1 + e_1 \\
3 f_- &=& 2p_1 + f_+ \;.
\eeq
This means that the conditions on $f_1$, $e_2$ and $f_-$ can be dropped.
Moreover, adding the fact that $\lambda_1, \lambda_2 >0$ is needed
for the potential to be bounded from below, we obtain
\beq
|c_+| &>& |c_-| \\
|b_+| &>& |b_-| \;.
\eeq
The resulting conditions for tree-level perturbative unitarity are thus
given by
\beq
|\lambda_3 - \lambda_4| &<& 8 \pi \label{eq:ev1} \nonumber \\
|\lambda_3 + 2 \lambda_4 \pm 3 \lambda_5| &<& 8 \pi \nonumber \\
\left| \frac{1}{2} \left( \lambda_1 + \lambda_2 + \sqrt{(\lambda_1 -
     \lambda_2)^2 + 4 \lambda_4^2}\right) \right| &<& 8\pi \nonumber \\
\left| \frac{1}{2} \left( \lambda_1 + \lambda_2 + \sqrt{(\lambda_1 -
     \lambda_2)^2 + 4 \lambda_5^2}\right) \right| &<& 8\pi. \label{eq:ev5}
\eeq
\end{itemize}

If a given choice of 2HDM parameters satisfies all of these constraints, it is accepted
(provided it further satisfies, at the weak scale, the experimental constraints described in the next section).
At this stage we include the effect of the renormalization group running of the parameters of
the theory to understand how it affects the allowed parameter space. We use the one-loop $\beta$-functions
for the parameters of the model (and also the VEVs $v_1$ and $v_2$), presented in Appendix~\ref{app:rge}, and
for each point in the parameter space chosen. We adopt the following procedure:
\begin{itemize}
\item Perform the RGE running of all potential parameters and VEVs starting at $m_Z$.
\item At each scale between $m_Z$ and the Planck scale, verify whether the theoretical
constraints detailed above (potential bounded from below; positive discriminant; perturbative
unitarity) are still verified.
\item If all the theoretical constraints are verified, proceed to a higher scale and repeat.
\item If at a given scale $\Lambda$ any of the theoretical constraints is not verified, stop the
RGE running and keep the information on this
cut-off scale.
\end{itemize}
There is a further constraint which must be considered -- for a large
region of the initial
parameter space, the RGE running will hit Landau poles -- {\em i.e.}, the parameters will tend to
infinity -- at some scale between $m_Z$ and the Planck scale. As in the SM, this is easily understood
if one considers the structure of the couplings in the $\beta$-functions of the model. For instance,
the contributions of the Yukawa couplings to the $\beta$-functions of
the quartic couplings are negative
and tend to reduce their values as one increases the renormalization scale; but the quartic couplings
have positive contributions to those $\beta$-functions and thus tend to increase their values. As a consequence,
only initial values of the quartic couplings with small magnitudes will not develop Landau poles during
the RGE running up to the Planck scale. For completion, we assume a Landau pole occurs if either
a) one of the gauge couplings, Yukawa couplings or quartic couplings of the potential reaches the absolute
 value of
100; b) if either $m_{11}^2$, $m_{22}^2$ or $m_{12}^2$ reaches the absolute
value of $10^{10} \times v^2 (v=246 \mbox{ GeV})$
or c) if $v_2/v_1 > 100$. Notice that since the $\beta$-functions are highly coupled, as soon as one
given parameter hits a Landau pole typically others will as well.

If some choice of parameters is such that one theoretical constraint is violated or a Landau pole occurs
at a given scale $\Lambda$, this means that the theory ceases to be valid above $\Lambda$ and requires
new physics (NP) above $\Lambda$ to correct the RGE evolution (for example, extra scalars to stabilize the
vacuum, or extra fermions to prevent Landau poles). Thus, if one believes that the 2HDM should be valid up
to a given high energy scale $\Lambda_{NP}$, the RGE running described will discard many combinations of
parameters, reducing the parameter space of the model and improving its predictability. The higher $\Lambda_{NP}$
is the more severe is the elimination of parameters. To give the reader an idea of the importance of each of
our requirements on the curtailment of the 2HDM parameter space, we found that the appearance of a Landau pole at
$\Lambda_{NP} = 1$ TeV reduces the number of original points to about 46 \%  in model type I, and 33 \%
in model type II\footnote{The $\beta$-functions are of course different for each of the model types considered,
see Appendix~\ref{app:rge}.};  by requiring the potential to also be bounded from below and unitarity to be
obeyed up to the same scale of 1 TeV will amount to a further reduction to 17 \% (8 \%) for type I (type II)
of the original points. Finally, the discriminant plays a very small role, with a further reduction of less than
1\% for again a scale of 1 TeV, for both model types. In fact, the discriminant will almost play no role in this
analysis --- the number of points which do not survive RGE running all the way to the Planck scale because
{\em only} the discriminant condition is violated is extremely
small.

Finally, a word on thresholds: we have taken the weak scale, $m_Z$, as the starting point of our RG analysis.
A more refined analysis would take into account the possibility of thresholds in the RG running
(for instance, using the 5-flavour $\beta$-functions between $m_Z$ and
the top threshold). Alternatively, we
could have started the RG running at a higher scale. In either case, the impact of these refinements in the
RG running in our analysis is minimal, at most slightly shifting the
cut off scales $\Lambda$. The substance of our conclusions would not be affected.

%%%%%%%%%%%%%%%%%%%%%%%%%%%%%%%%%%%%%%%%%%%%%%%%%%%%%%%
\section{The 2HDM parameter space \label{sec:2hdmspace}}

The 2HDM is implemented as a model class in {\tt ScannerS}~\cite{Coimbra:2013qq,ScannerS}, and we used
this code to generate our data samples. The
theoretical bounds described in the previous section, plus all relevant available
experimental constraints, are either inbuilt in the code or interfaces with several other
codes allow to take them into account in the sample generation.

We will now briefly describe the experimental constraints on the model and how
they are applied. The most relevant exclusion bounds on the $m_{H^\pm}-t_\beta$ plane are those
which arise from the $B \to X_s \gamma$  measurements~\cite{Deschamps:2009rh,Mahmoudi:2009zx,Hermann:2012fc,
Misiak:2015xwa,Misiak:2017bgg}.
A $2\sigma$ bound on the charged Higgs mass of $m_{H^\pm} > 580 \mbox{
  GeV}$ for the models type II and Flipped
that is almost independent of $\tan \beta$ was recently discussed in~\cite{Misiak:2017bgg}.
In all types of 2HDMs, there is also a hard bound on the charged Higgs mass coming
from LEP, with the process $e^+ e^- \to H^+ H^-$~\cite{Abbiendi:2013hk} which is approximately
$100  \mbox{ GeV}$. We have used all the flavour constraints, plus the ones from
the $R_b$~\cite{Haber:1999zh,Deschamps:2009rh} measurement.
These constraints are $2\sigma$ exclusion bounds on the $m_{H^\pm}-t_\beta$ plane.
Furthermore, all points comply with the electroweak precision measurements. We demand a
$2\sigma$ compatibility of the $S$, $T$ and $U$ parameters with the SM fit presented in~\cite{Baak:2014ora}.
The full correlation among these parameters is taken into account.

The mass of the SM-like Higgs boson, denoted by $h_{125}$, is set to
$m_{h_{125}} = 125.09 \; \mbox{GeV}$
\cite{Aad:2015zhl}.
The {\tt HiggsBounds} code
  \cite{Bechtle:2008jh,Bechtle:2011sb,Bechtle:2013wla} is used to check for agreement
with all 2$\sigma$ exclusion limits from LEP, Tevatron and LHC Higgs searches.
The decay widths and branching ratios were calculated with {\tt HDECAY}
\cite{Djouadi:1997yw,Butterworth:2010ym}, which includes off-shell
decays and state-of-the-art QCD corrections.
All Higgs boson production cross sections via gluon fusion ($ggF$) and $b$-quark
fusion ($bbF$) are obtained from {\tt SusHiv1.6.0}
\cite{Harlander:2012pb,Harlander:2016hcx},
at NNLO QCD. The SM-like Higgs rates are forced to be within $2\times 1\sigma$
of the fitted experimental values given in~\cite{Khachatryan:2016vau}. In that
reference bounds are presented for the quantities
\beq
\frac{\mu_F}{\mu_V} \;, \quad \mu_{\gamma\gamma} \;, \quad
\mu_{ZZ} \;, \quad \mu_{WW} \;, \quad  \mu_{\tau\tau} \;, \quad
\mu_{bb} \;,
\eeq
where $\mu_F$ ($\mu_V$) is the ratio, for each channel, between the measured
cross section, and its SM expected value, for the gluon-gluon fusion and $t\bar{t}H$
(VBF + VH) production processes; the quantities $\mu_{xx}$ are then defined as
\beq
\mu_{xx} = \mu_F \, \frac{\mbox{BR}_{\text{2HDM}} (H_i \to
  xx)}{\mbox{BR}_{\text{SM}} (H_{\text{SM}} \to xx)}
\eeq
for $H_i \equiv h_{125}$ and the SM Higgs boson $H_{\text{SM}}$. Because custodial symmetry is preserved,
$\mu_{ZZ} = \mu_{WW} \equiv \mu_{VV}$, and we are allowed to combine the lower
$2 \times 1\sigma$ bound from $\mu_{ZZ}$ with the upper bound on $\mu_{WW}$~\cite{Khachatryan:2016vau},
\beq
0.79 < \mu_{VV} < 1.48 \;.
\eeq

In type II we choose the charged Higgs mass to be in the range
\beq
580 \mbox{ GeV } \le m_{H^\pm} < 1 \mbox{ TeV } \;,
\eeq
while in type I we have taken
\beq
80 \mbox{ GeV } \le m_{H^\pm} < 1 \mbox{ TeV } \;.
\eeq
Taking into account all the constraints, in order to optimise
the scan, we have chosen the following regions in the remaining input parameters:
$0.8 \le \tan \beta  \le 35$, $- \frac{\pi}{2} \le \alpha < \frac{\pi}{2}$,
$ 0 \mbox{ GeV}^2 \le m_{12}^2  < 500 000 \mbox{ GeV}^2$, $30 \mbox{ GeV} \le m_{A}  < 1000 \mbox{ GeV}$
and finally $130 \le m_H < 1000 \mbox{ GeV}$ for the
  light Higgs scenario, but $30 < m_h < 120$~GeV for the heavy Higgs scenario.

\section{The light Higgs scenario \label{sec:res1}}

In this section we will consider the standard approach to the 2HDM, in which the lightest
CP-even scalar is taken to be the observed 125 GeV Higgs boson. The heavy Higgs scenario is
dealt with in the next session. Our goal now is to carefully analyse
what impact the requirement
of imposing the theoretical constraints described in the previous
section has, plus the absence of Landau
poles, for all scales above the weak scale.

\subsection{Results with no collider bounds}

We start the discussion with a sample of points that have passed all the theoretical
constraints, the electroweak precision tests and all $B$-physics
constraints -- the most important one being
the constraint from $b \to s \gamma$, which in type II forces the charged Higgs mass to be above 580 GeV
at 2$\sigma$. However, we have not imposed the LHC bounds on the observed Higgs rates from
\cite{Khachatryan:2016vau} on this parameter sample. What we will observe is
that the requirement that the
potential is well-behaved for increasingly high energy scales will curtail the parameter space so much
that, in some situations, the 125 GeV scalar becomes ``naturally" aligned.

\begin{figure}[b!]
  \centering
  \includegraphics[width=0.47\linewidth]{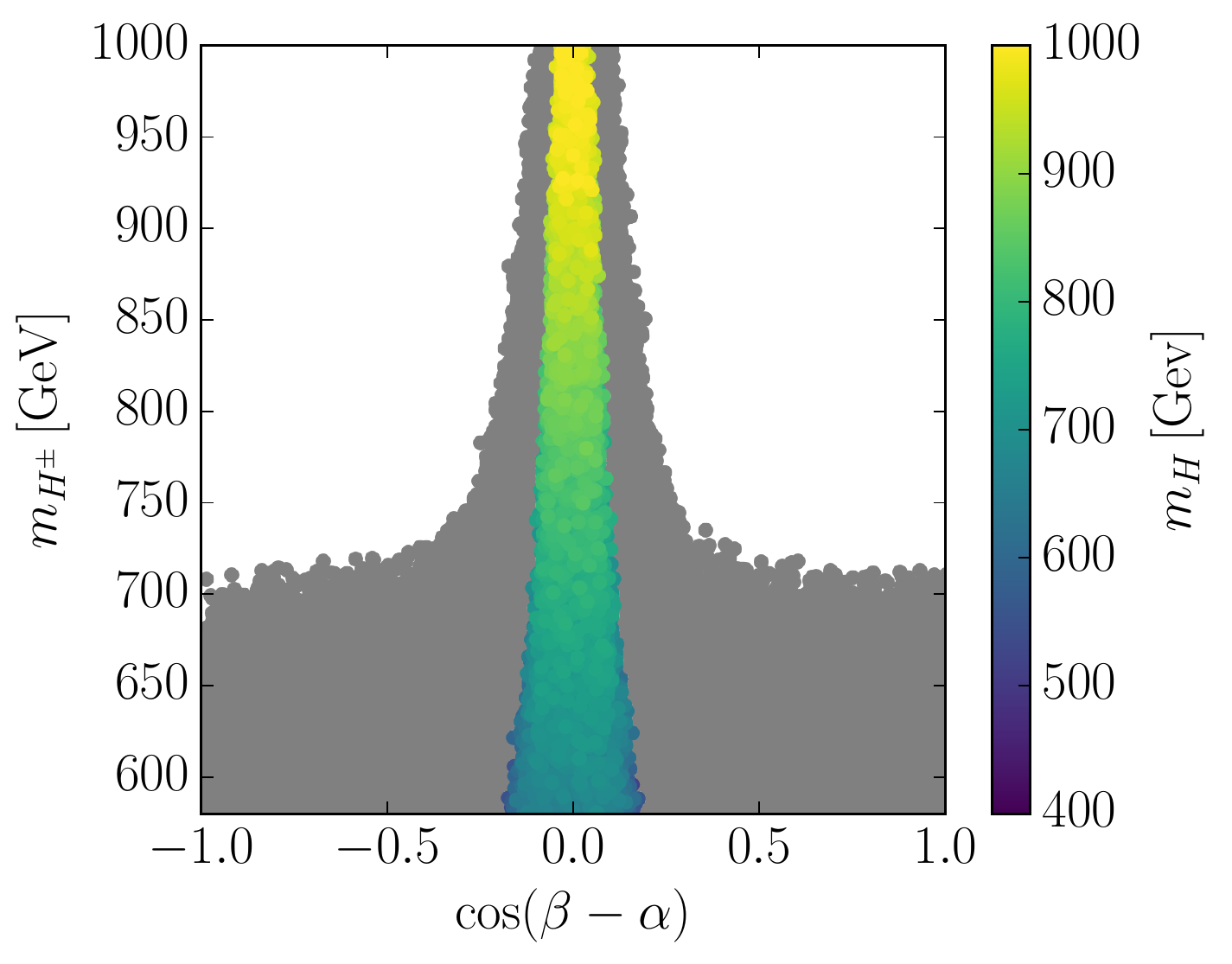}
  \includegraphics[width=0.47\linewidth]{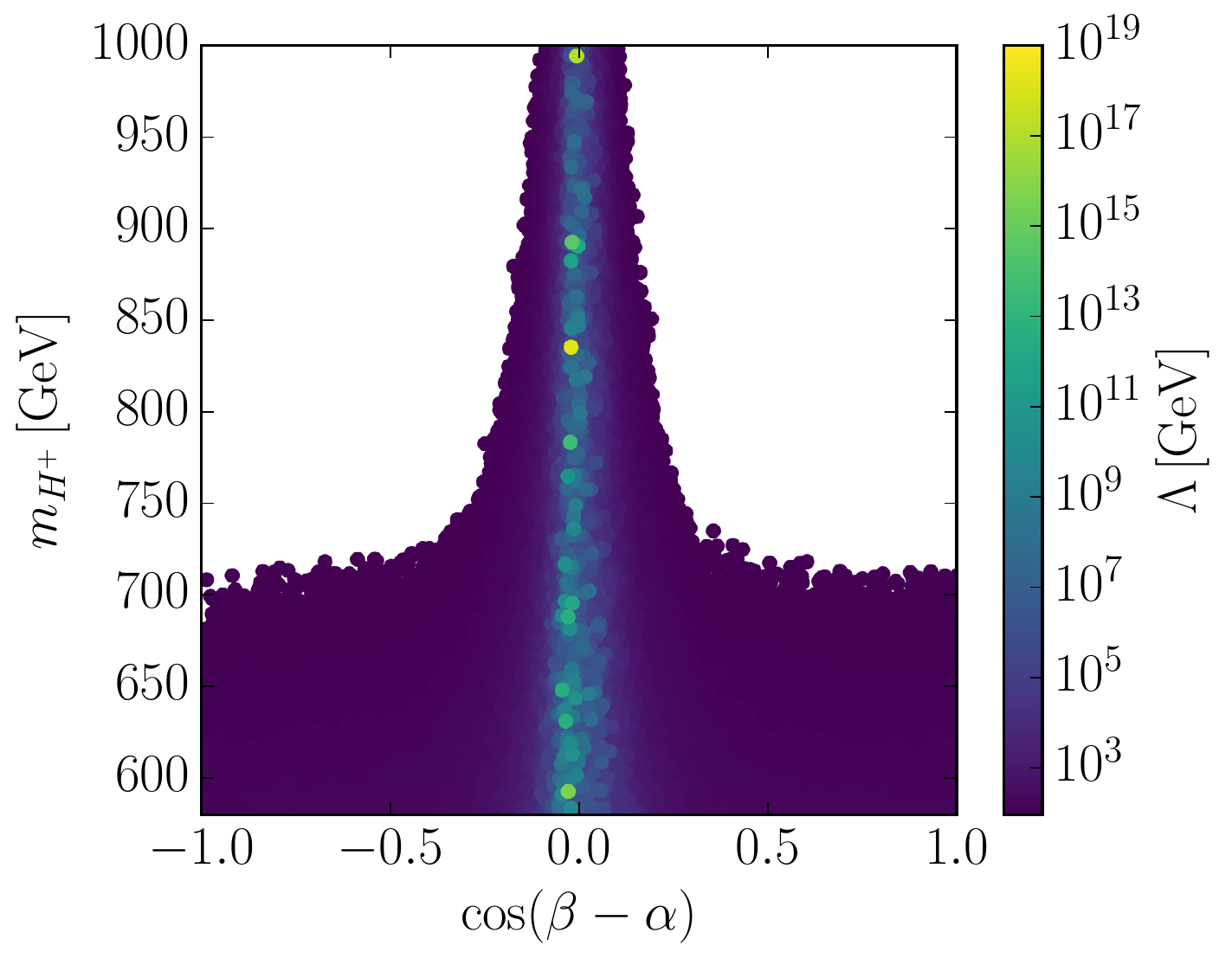}
  \caption{Charged Higgs mass vs. $\cos(\beta - \alpha)$ in the type II 2HDM. On the left panel we show, in grey,
   the points that passed the theoretical, electroweak and $B$-physics constraints  at the scale $m_Z$.
   The remaining points have survived the RGE running up to a scale of 1 TeV. The colour bar shows the value of $m_H$. On the right panel we present the same plot but where the colour
   bar shows the cut-off scale. The points are sorted from dark to brighter colours. }
 \label{fig:MHC-CosBA_II_no_bounds}
\end{figure}

Let us begin with the analysis of the type II model. The data sample we used had almost 1 million
different parameter combinations, and for each of those points we performed the RGE running
described above, verifying the cut-off scales $\Lambda$ for which either the theoretical constraints
we imposed were violated or a Landau pole was reached. The results of this work allowed us
to obtain Fig.~\ref{fig:MHC-CosBA_II_no_bounds}, which we now analyse in detail.

On the left panel of Fig.~\ref{fig:MHC-CosBA_II_no_bounds} we present the
charged Higgs mass vs.~$\cos(\beta - \alpha)$. We show in grey
the points that passed the theoretical, $S,T,U$ and $B$-physics constraints.
Notice how clearly the LHC bounds were not present in the initial sample, since
$\cos (\beta -\alpha)$ varies from -1 to 1, whereas current experimental results point to
the observed Higgs having SM-like behaviour, which would necessitate
values of $|\cos (\beta -\alpha)|$ much closer to zero. The coloured
points in the plot are the subset of the initial data
sample which survived the RGE running up to a scale of 1 TeV -- meaning, for which no theoretical constraint
was violated, nor a Landau pole occurred, between $m_Z$ and 1 TeV. The colour bar shows the minimum value of $m_H$,
 which shows a similar trend to the charged Higgs mass. For example, with the cutoff
scale of 1 TeV, the minimum value of $m_H$ allowed is  about 440 GeV . It is clear
that, already at such a small scale as 1 TeV, the range of variation of $\cos (\beta -\alpha)$ has
shrunk from the original $|\cos (\beta -\alpha)| < 1$ to about $|\cos (\beta -\alpha)| < 0.2$ --- {\em which
means that the simple requirement that the type II potential is well behaved up to a scale of about
1 TeV implies that the Higgs boson must have SM-like behaviour} provided that one of the scalars
has a mass above $\approx$ 500 GeV. Here we see alignment arising in a
``natural" way from the behaviour of the theory, rather than requiring
a particular choice of the parameter
region to fit the data.

On the right panel of Fig.~\ref{fig:MHC-CosBA_II_no_bounds} we show the result of continuing the
RGE running for higher scales than 1 TeV. Again we plot $m_{H^\pm}$ vs~$\cos(\beta - \alpha)$, but
now the colour code provides information on the cut-off scale $\Lambda$, that is,
the scale at which either a Landau pole occurs or any of the theoretical conditions is violated. As
expected from previous analyses, many points survive up to the Planck scale with values of
$\cos (\beta - \alpha)$ increasingly closer to zero.  The 2HDM of type II can therefore be a valid
description of particle physics all the way up to the Planck scale -- and since we made sure the potential
is bounded from below and the correct electroweak minimum is the global one at all scales, we conclude
that within the framework of this model it is possible to choose parameters to even avoid the issue of
metastability which has been discussed for the SM.
\begin{figure}[b!]
  \centering
  \includegraphics[width=0.4\linewidth]{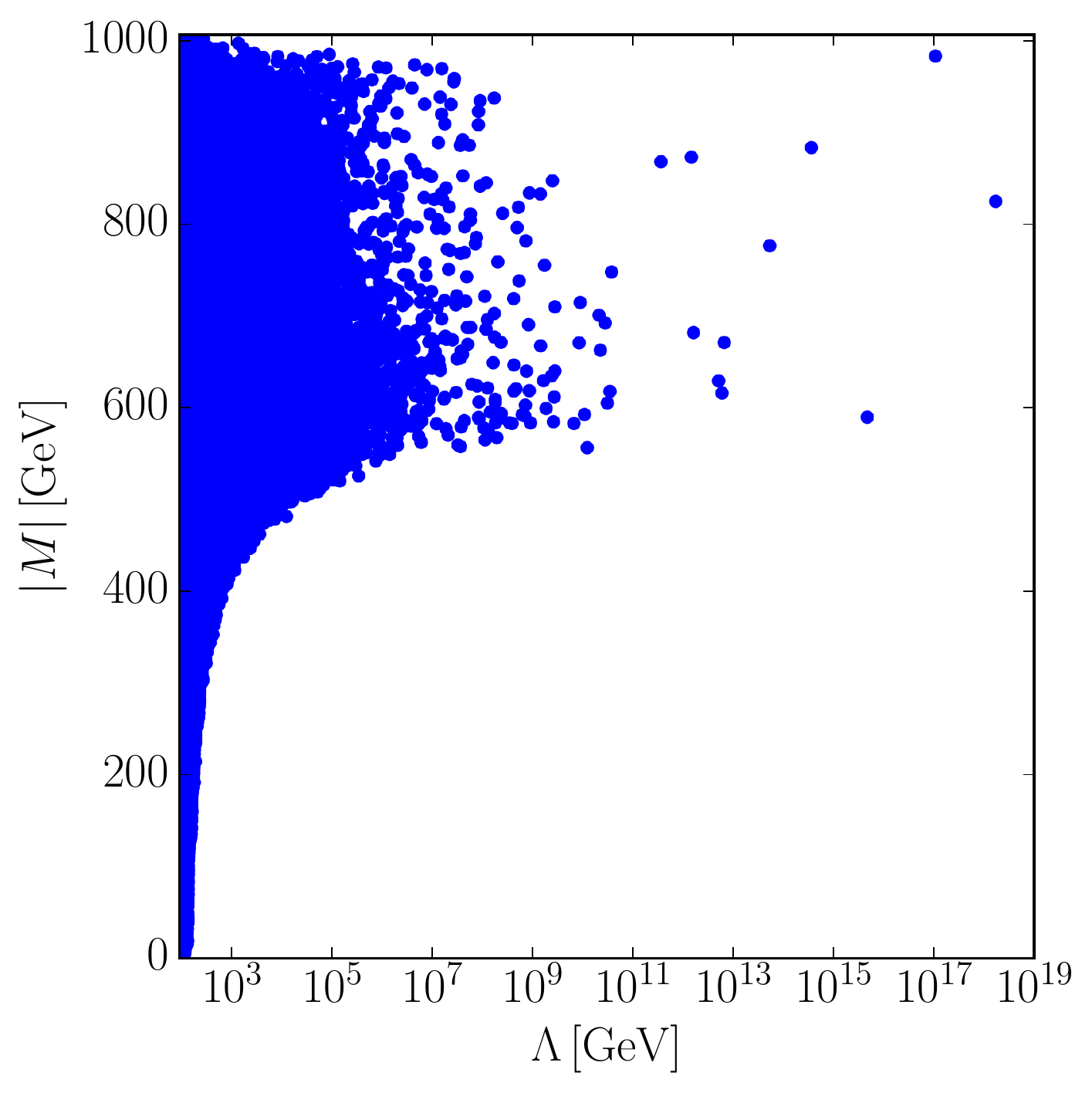}
  \includegraphics[width=0.4\linewidth]{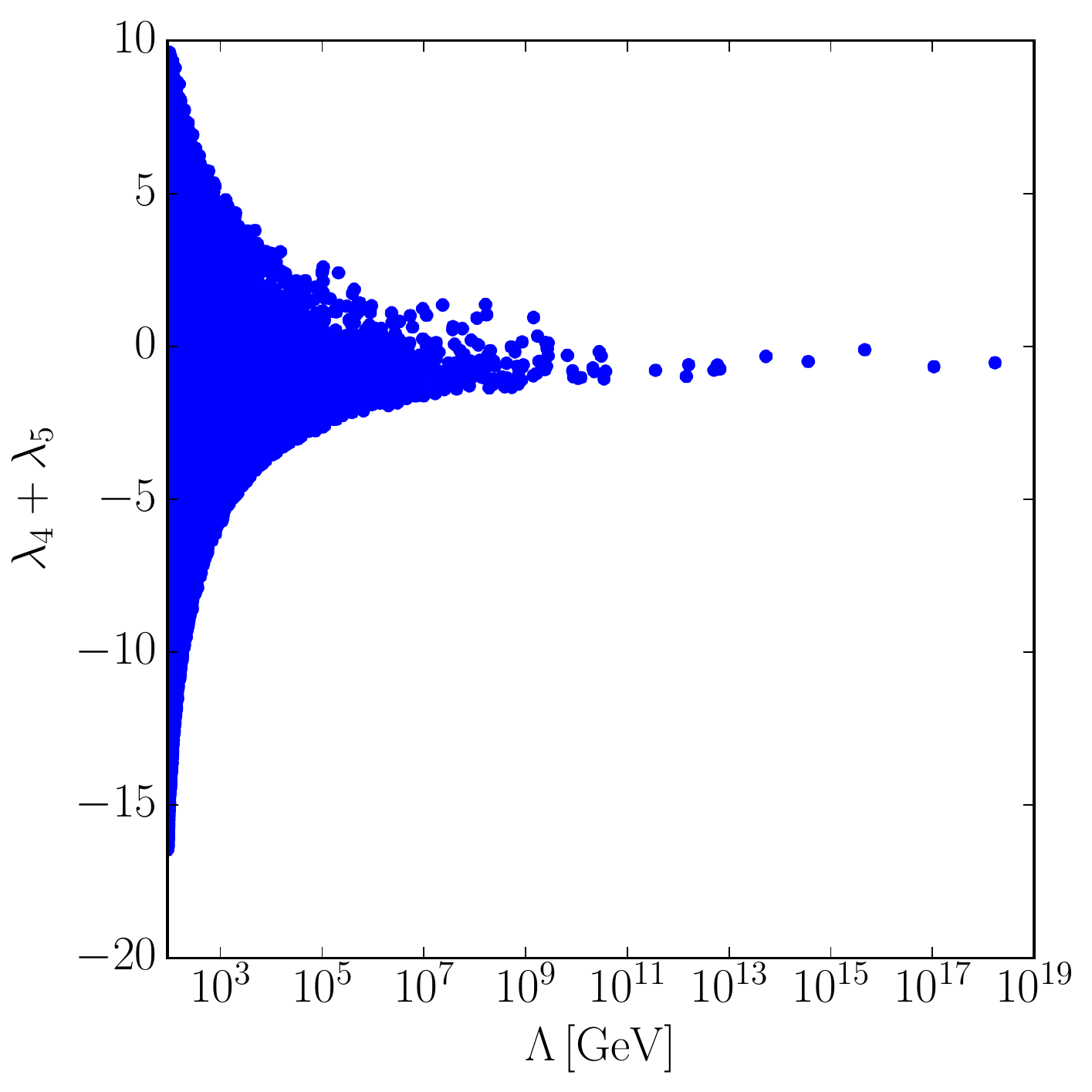}
  \caption{On the left (right) panel we present $|M|$ ($\lambda_4 + \lambda_5$) as a function of
  the cut-off scale $\Lambda$ in the type II 2HDM. The points
  have passed both the theoretical constraints and $b \to s \gamma$ at the scale $m_Z$ and
  have also survived the RGE running up to the cut-off scale.
    }\label{fig:explainmassdiff_no_bounds}
\end{figure}
\begin{figure}[h!]
  \centering
  \includegraphics[width=0.47\linewidth]{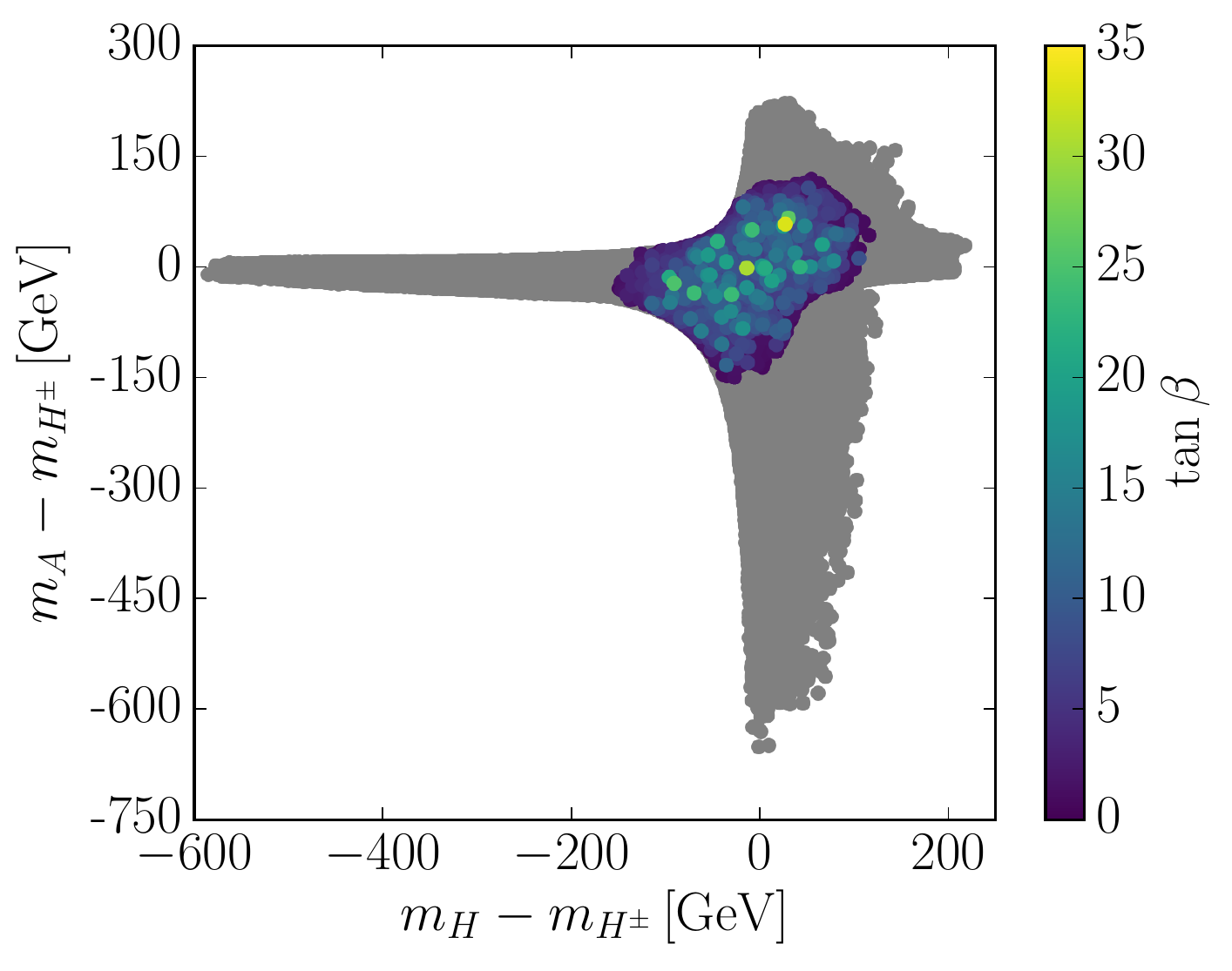}
  \includegraphics[width=0.47\linewidth]{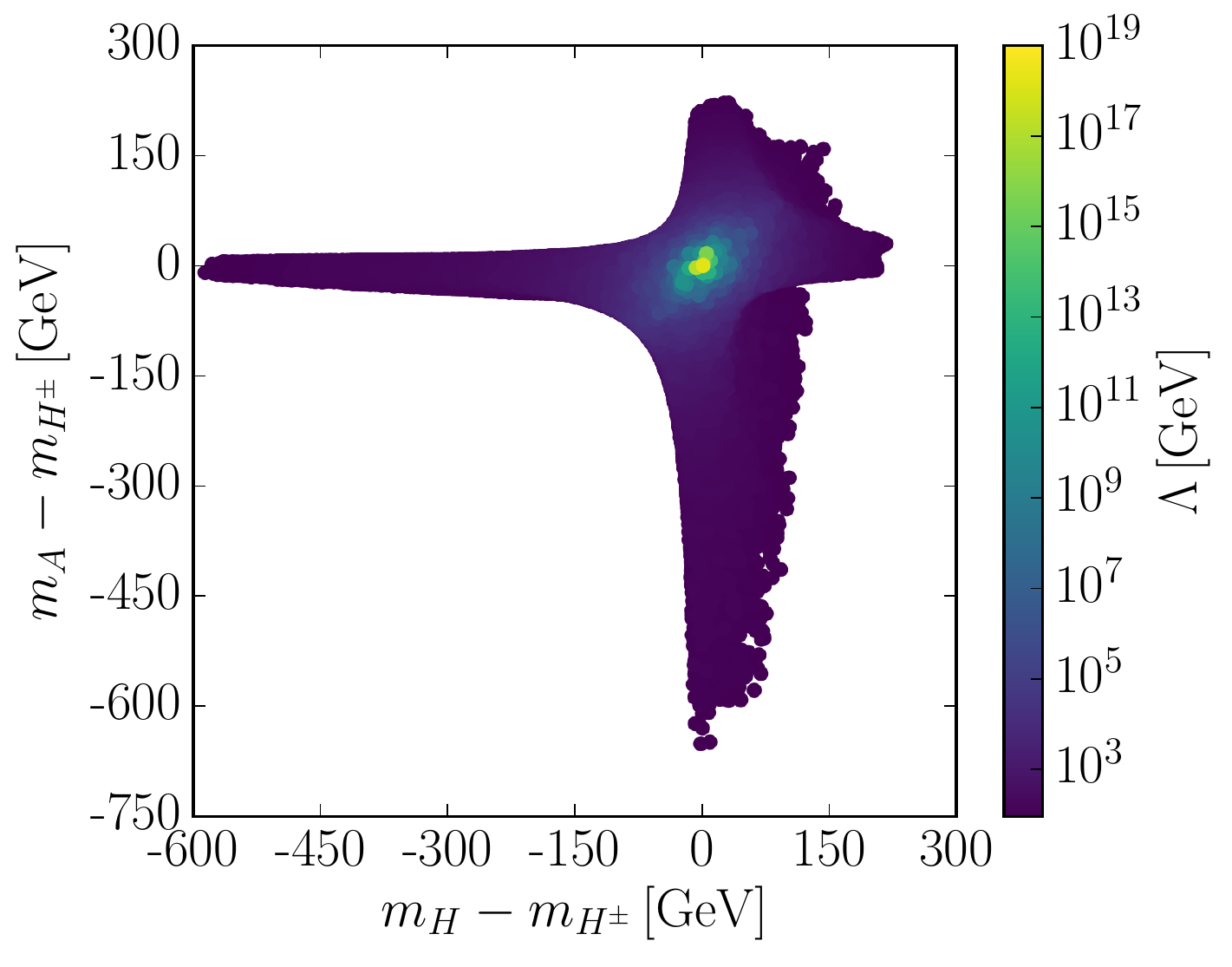}
  \caption{$m_A - m_{H^\pm}$ vs. $m_H - m_{H^\pm}$ in the type II 2HDM. On the left panel we show in grey
   the points that passed the theoretical, electroweak and $B$-physics constraints at the scale $m_Z$.
   The remaining coloured points survived the RGE running up to a scale of 1 TeV. The colour bar shows
   the values of $\tan \beta$.
   On the right panel we present the same plot but where the colour
   bar shows the cut-off scale.
    }\label{fig:massdiff_II_no_bounds}
\end{figure}
What is the origin of the quick approach of the alignment regime, at such a
remarkably low energy scale as 1 TeV? In Fig.~\ref{fig:explainmassdiff_no_bounds} we present
the points that have survived the running up to the cut-off scale $\Lambda$. On the left we see
that $M$ has a very fast increase, reaching a maximum value of $~423$ GeV at 1 TeV. For higher scales we see
that $M$ can take many values, but its minimum value then stabilizes, with a value close to the charged Higgs mass
at $\Lambda \approx 10^3$ TeV, and remains constant up to the Planck scale.

On the right panel of Fig.~\ref{fig:explainmassdiff_no_bounds} we show the
values of $\lambda_4 + \lambda_5$ for the points which survive from $m_Z$ up to the Planck scale.
Clearly, the absolute value of $\lambda_4 + \lambda_5$ is decreasing, and the sum of the two couplings
can only take values close to zero if the model is to be valid up to very high energy scales. Indeed,
already at 1 TeV we have $|\lambda_4 + \lambda_5| \lesssim 5.7$ and we can attempt an analytical explanation
for the approach to alignment, using
\begin{equation}
M^2 - m_{H^\pm}^2= - \frac{\lambda_4 + \lambda_5}{2} v^2 \quad
\Rightarrow \quad M - m_{H^\pm}
= - \frac{(\lambda_4 + \lambda_5) v^2}{2 (M + m_{H^\pm})} \;.
\end{equation}
Inserting the maximum value for $\lambda_4 + \lambda_5$ and the minimum value for $m_{H^\pm}$ and $M$
for 1 TeV in these formulae, we obtain $|M - m_{H^\pm}| \approx 163$ GeV.
This provides an approximation for the maximum
mass difference between the several scalars. To reinforce this point, in
Fig.~\ref{fig:massdiff_II_no_bounds} we plot $m_A - m_{H^\pm}$ vs. $m_H - m_{H^\pm}$ in the type II 2HDM.
Again, the grey points have passed all theoretical constraints and comply with $b \to s \gamma$ at the scale
$m_Z$. On the left plot, the remaining points, colour coded with the values of $\tan \beta$, have survived
the RGE running up to a scale of 1 TeV. As discussed, we can see clearly in the plot
that all mass differences are below $\pm$200~GeV -- and since in type II the charged Higgs mass is constrained
by the $b \to s \gamma$ results to be above 580 GeV, this gives us possible lower bounds on the masses for
the pseudoscalar or the heavier CP-even scalar of about 430 GeV.

Let us now consider the right plot of Fig.~\ref{fig:massdiff_II_no_bounds},
where we can see the dependence of the cut-off scale. Because the absolute values of the quartic couplings
decrease and the lowest value of $M$ stabilizes with   increasing cut-off scale,
the mass differences approach zero. As the quartic couplings decrease, then, all masses are
increasingly controlled by $M$ and they tend to be of the same order. Thus, validity at very high scales
($\gg$ 1 TeV) implies, for model type II, that all extra scalars have necessarily high masses, and alignment
is reached via decoupling -- we found the minimum acceptable values for the scalar masses if the model is valid
above $10^{11}$ GeV is roughly 600 GeV. However, if one is more conservative and only assumes that the 2HDM
describes particle physics up to a scale of about 1 TeV, then alignment can be reached with relatively low
masses (450 GeV) for $A$ and $H$, and one could argue that we are
observing the alignment limit, instead of the decoupling limit. In either case,
though, the result is the same: requiring the type II 2HDM to be well behaved at least up to a scale of 1 TeV
automatically means that alignment must be satisfied.

\begin{figure}[t!]
  \centering
  \includegraphics[width=0.47\linewidth]{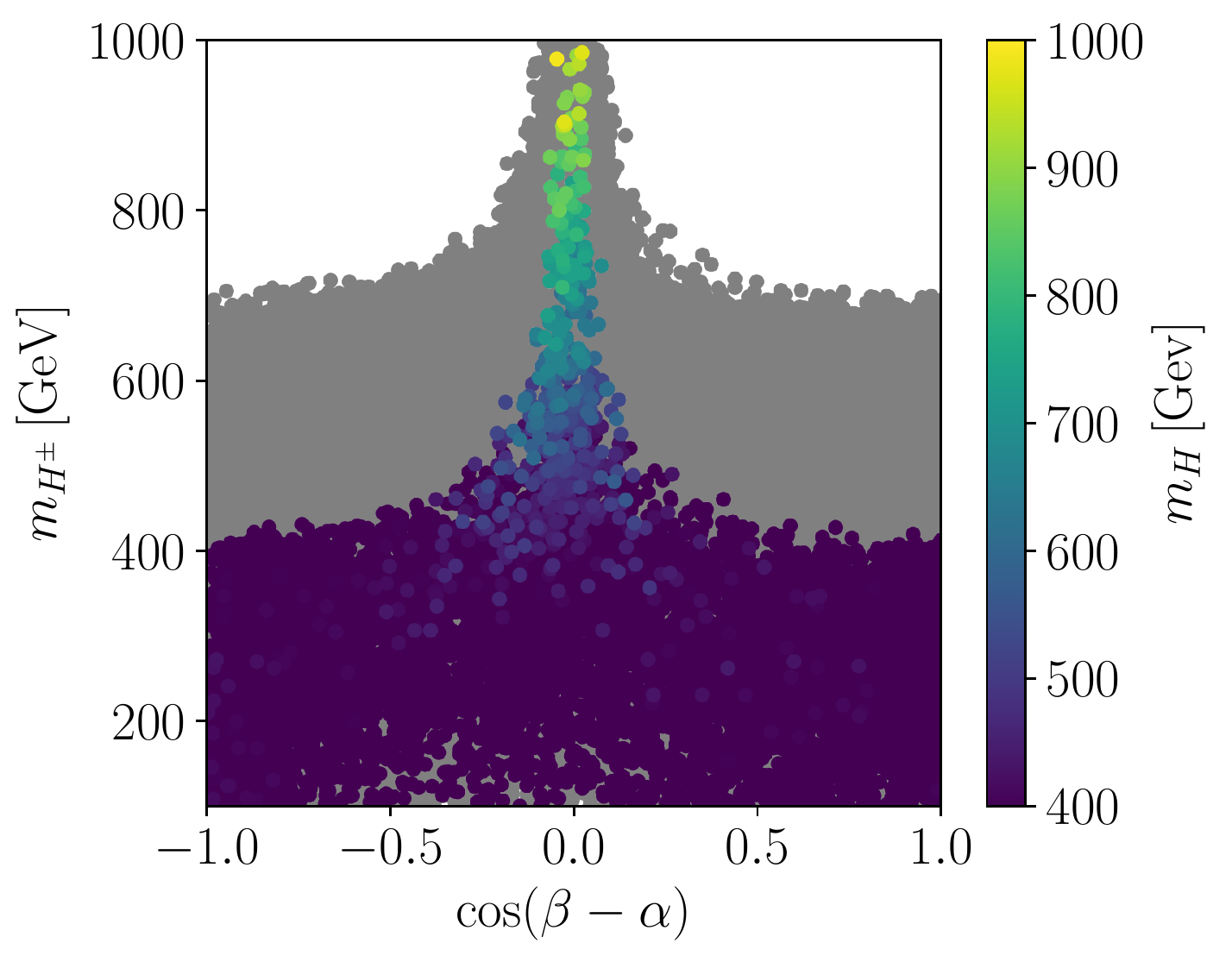}
  \includegraphics[width=0.47\linewidth]{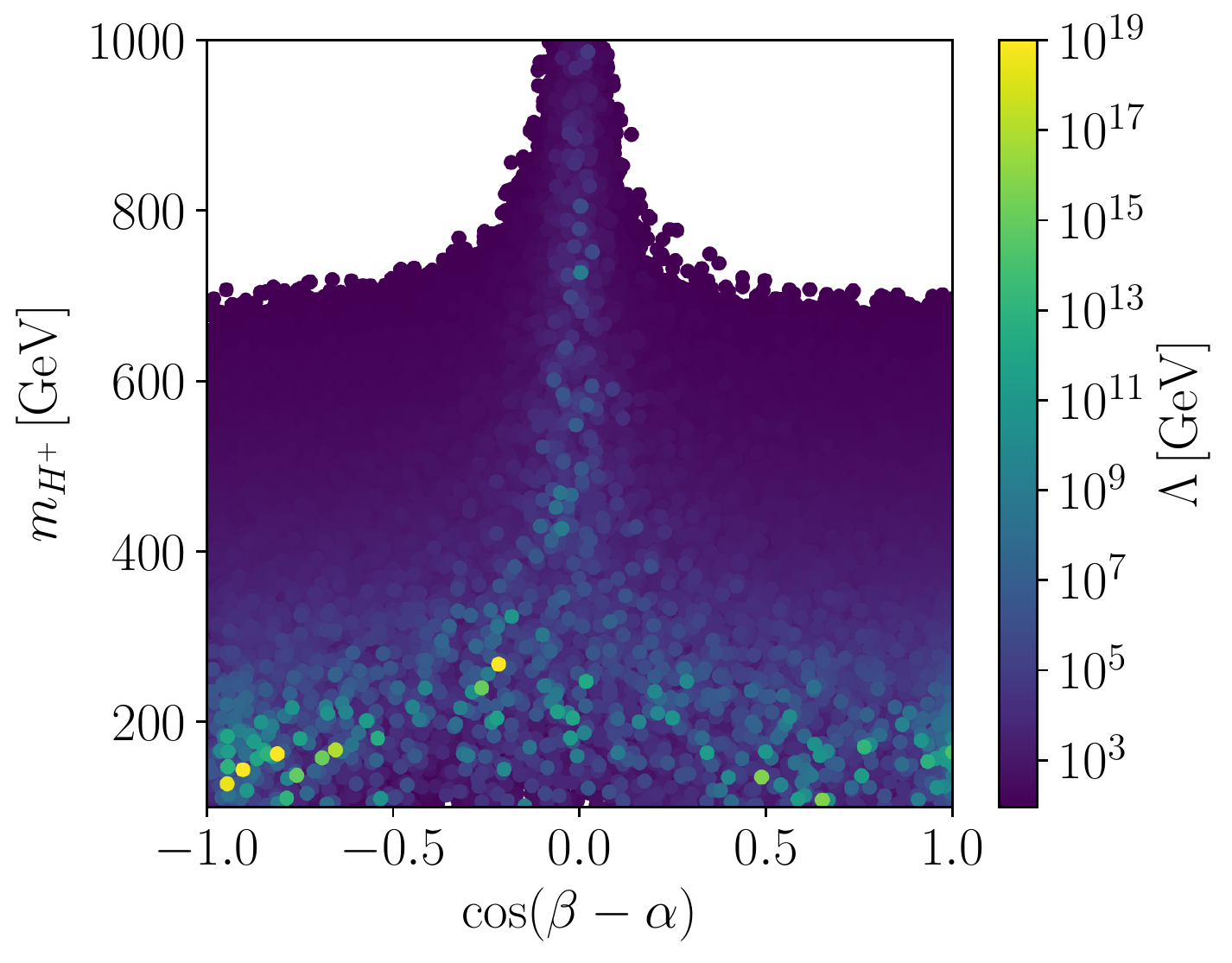}
  \caption{Charged Higgs mass vs. $\cos(\beta - \alpha)$ in the type I 2HDM. On the left panel we show, in grey,
   the points that passed the theoretical, $S, T, U$ and $B$-physics constraints at the scale $m_Z$.
   The remaining points have survived the RGE running up to a scale of 1 TeV. The colour bar shows the value of $m_H$. On the right panel we present the same plot but where the colour
   bar shows the cut-off scale. }
 \label{fig:MHC-CosBA_I_no_bounds}
\end{figure}

The situation is, however, different for a 2HDM of type I. In Fig.~\ref{fig:MHC-CosBA_I_no_bounds} we
show the analog, for type I, of
Fig.~\ref{fig:MHC-CosBA_II_no_bounds}. And the striking difference is
that requiring the validity of the model up to high energy scales does not
necessarily imply alignment for the lightest Higgs boson --
we see, in the plot of the right, plenty of points away from alignment (with large absolute values of
$\cos(\beta - \alpha)$) which
survive all the way up to the Planck scale. Thus validity of the 2HDM up to high energy scales does
not necessarily imply alignment for type I, though it does for type II. The left plot of
Fig.~\ref{fig:MHC-CosBA_I_no_bounds} shows that the charged Higgs mass
is playing a crucial role in this respect
-- in fact, if in model type I one were to impose $m_{H^\pm} > 500$ GeV, again one would have alignment emerging
from requiring that the model be valid up to energy scales of at least 1 TeV. However, for model type I there is
no compelling physics reason to impose such a cut on the charged mass, unlike what happens in type II. Still, the
reasoning can be inverted: if particle physics is described by a type I 2HDM, the fact that LHC indicates that
the lightest scalar is aligned means that, for the model to be valid up to very high energy scales, the
``natural" expectation is to have a charged Higgs mass superior to 500
GeV. This may be understood from the right plot in Fig.~\ref{fig:MHC-CosBA_I_no_bounds} -- points where validity 
occurs up to the Planck scale with lower charged masses are certainly possible, but not necessarily aligned.
Thus, if the charged mass is below roughly 500 GeV, validity of the model up to the Planck scale is possible,
but alignment does not arise ``naturally", it needs to be further imposed on the model, as a fine-tuning of its
parameters. If $m_{H^\pm} > 500$ GeV, on the other hand, validity up to scales as low as $\sim$ 1 TeV already
implies alignment.

\subsection{Results with collider bounds}

In the previous section we showed how alignment arises, in type II,
from requiring that the 2HDM be valid up to high energy scales. In
type I alignment does not arise automatically from that
requirement, unless one further demands that the charged Higgs mass be
superior to 500 GeV. Let us now see what the requirement of validity
to high energy scales combined with the LHC bounds on Higgs physics
teaches us about 2HDM phenomenology.
In this section we have used a sample where, besides all theoretical
constraints, electroweak precision bounds and  constraints from $B$-physics, we have imposed all
available collider bounds, and in particular those that restrict the
Higgs couplings to fermions and gauge bosons, which are the most
relevant ones.
The further constraints arising from requiring validity
of the model to high scales will obviously increase the predictability
of the theory.

A first observation to take into account is that LHC collider bounds
are a lot less restrictive for the 2HDM type I than for type II (see,
for instance,~\cite{Ferreira:2011aa,Ferreira:2014dya}). Even
with the latest run II data, the allowed parameter space permits
substantial deviations from alignment.
\begin{figure}[t!]
  \centering
  \includegraphics[width=0.47\linewidth]{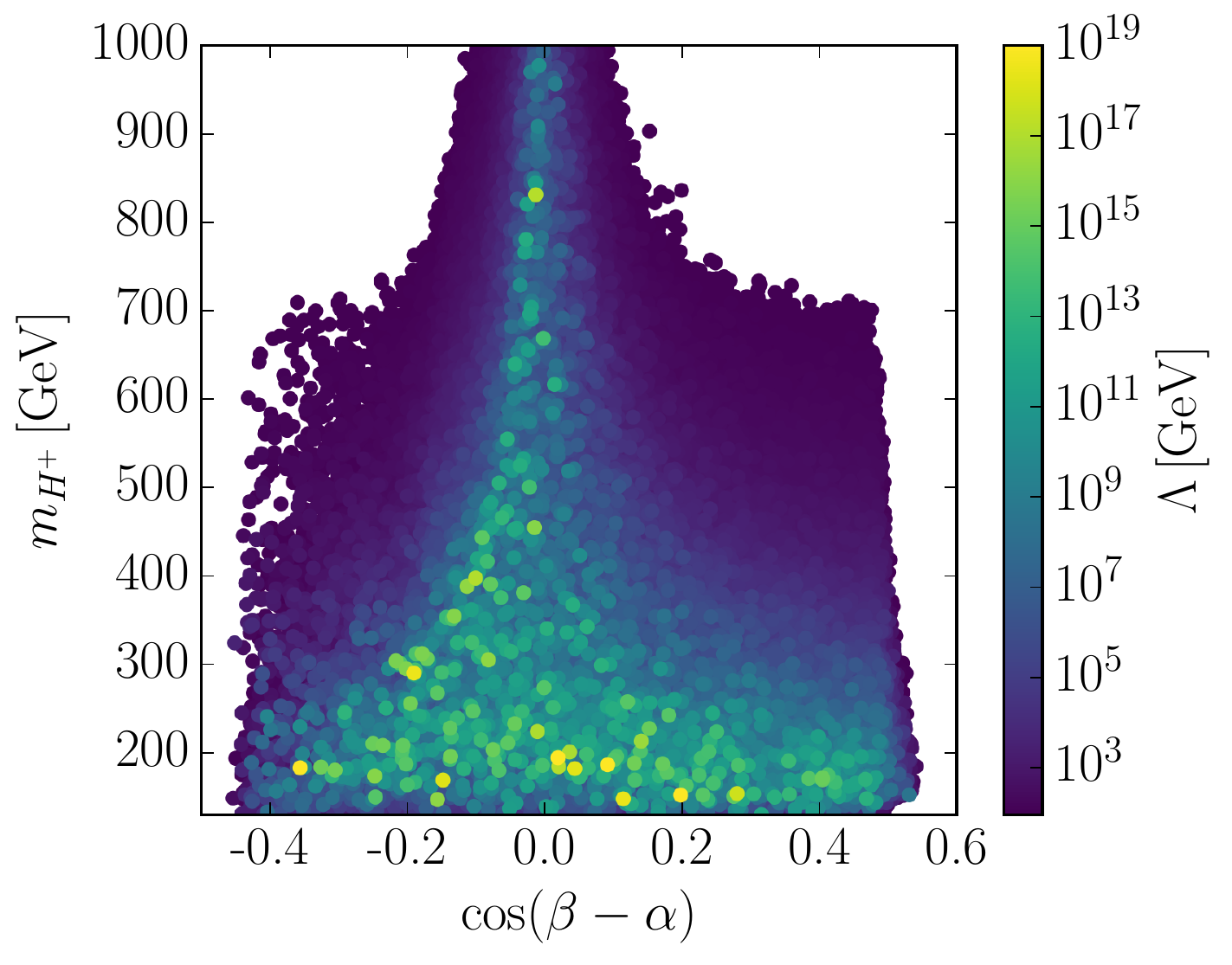}
  \includegraphics[width=0.47\linewidth]{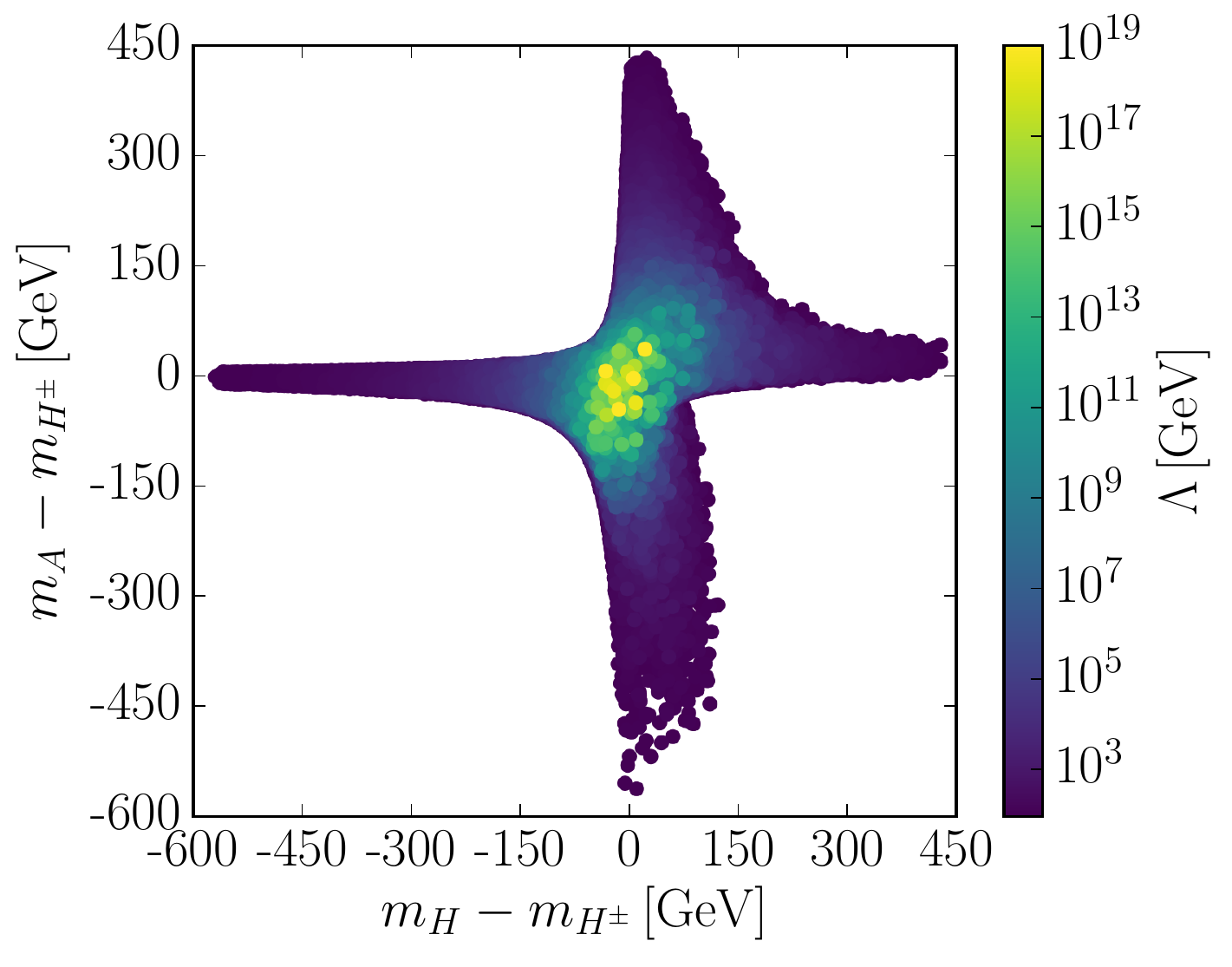}
  \caption{On the left panel we present the charged Higgs mass
    vs. $\cos(\beta - \alpha)$ in the type I 2HDM,
  colour coded with the cut-off scale. The points have passed all constraints at the scale $m_Z$ and have
  survived up to a given cut-off scale. On the right we present the plot for the mass differences
  $m_A - m_{H^\pm}$ vs. $m_H - m_{H^\pm}$ as a function of the cut-off scale.
    }\label{fig:MHC-CosBA_I}
\end{figure}
We start by presenting in Fig.~\ref{fig:MHC-CosBA_I} (left) a plot for the
type I 2HDM in the charged Higgs mass vs. $\cos(\beta - \alpha)$ plane. As previously discussed, because there
are no strong bounds on the charged Higgs mass, nor on any other scalar besides the 125 GeV one,
there is no major difference in the allowed range of $\cos(\beta - \alpha)$
for low and high scales of validity of the theory. This means that,
whatever the collider bounds on the type I models are, the model may be valid up to the Planck
scale even with large deviations from the alignment limit. Notice the yellow points in the left
plot of Fig.~\ref{fig:MHC-CosBA_I} with charged Higgs masses as low as $\sim$ 150 GeV and large absolute values
of $\cos(\beta - \alpha)$, corresponding to 2HDM type I parameter sets for which the model is valid up to
the Planck scale --- and while not satisfying alignment, they still satisfy all existing LHC bounds.

\begin{figure}[ht!]
  \centering
  \includegraphics[width=0.4\linewidth]{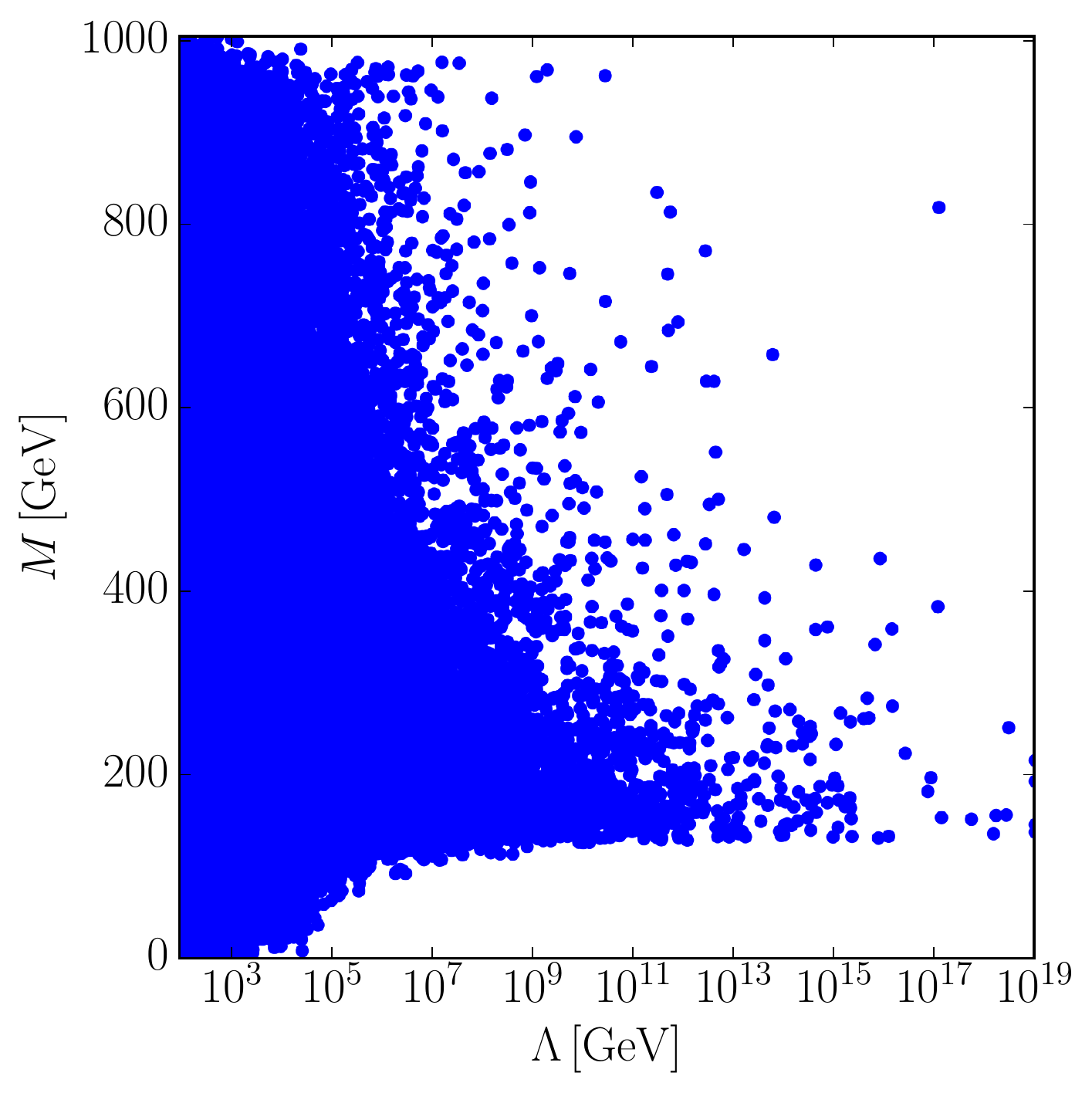}
  \caption{$M$ as a function of the cut-off scale $\Lambda$ for type
    I. These points fulfil all experimental and theoretical bounds.
    }\label{fig:symZ2}
\end{figure}

Still, we would recover the type II results if we had the same bound on the charged Higgs mass as for type II
-- once again, if the charged mass is superior to 500 GeV, alignment is an automatic consequence of
requiring validity of the model up to high scales.
On the right plot we present $m_A - m_{H^\pm}$ vs. $m_H - m_{H^\pm}$
as a function of the cut-off scale. As for the type II model, the mass
differences become increasingly smaller with increasing $\Lambda$,
which again suggests that they are controlled by the scale $M$ especially when we move closer to the Planck scale where the quartic couplings
are extremely small in magnitude. The reason why we are not driven to
decoupling in the generic type I is shown in Fig.~\ref{fig:symZ2}.
%
%\ref{fig:Z6Z1_I}.
Contrary to type II, and although the value of $M$ increases also very
fast, its lowest value stabilizes at a value close to 100 GeV. Since
for high values of the cut-off scale $\Lambda$ the quartic couplings
are very close to zero, their contribution to the masses is negligible
when compared to that of $M$ (see \eqref{eq:masses}). Thus,
light masses, of order 100--200 GeV, are still allowed even   if the
type I model is valid up to high scales.

\begin{figure}[b!]
  \centering
  \includegraphics[width=0.47\linewidth]{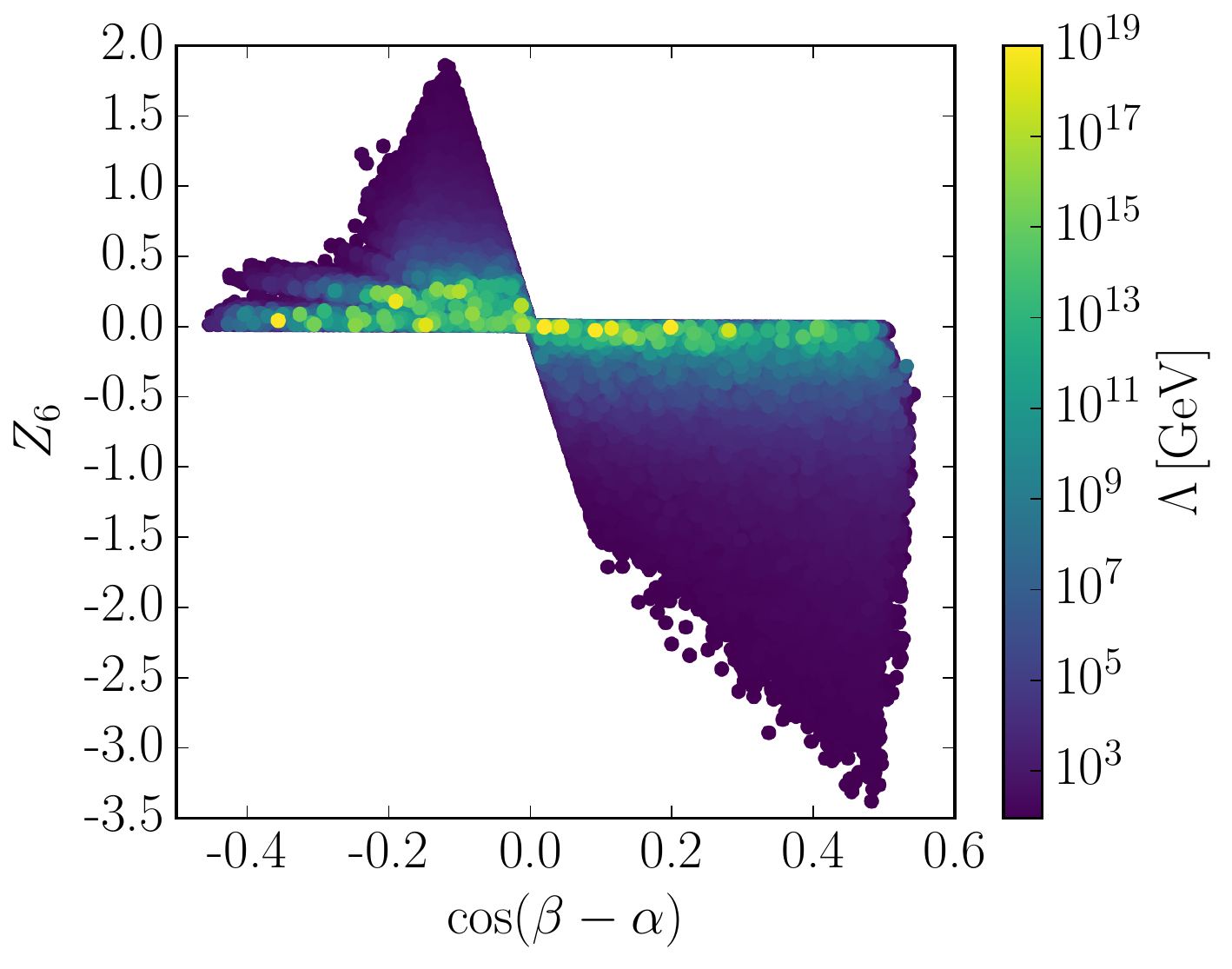}
  \includegraphics[width=0.47\linewidth]{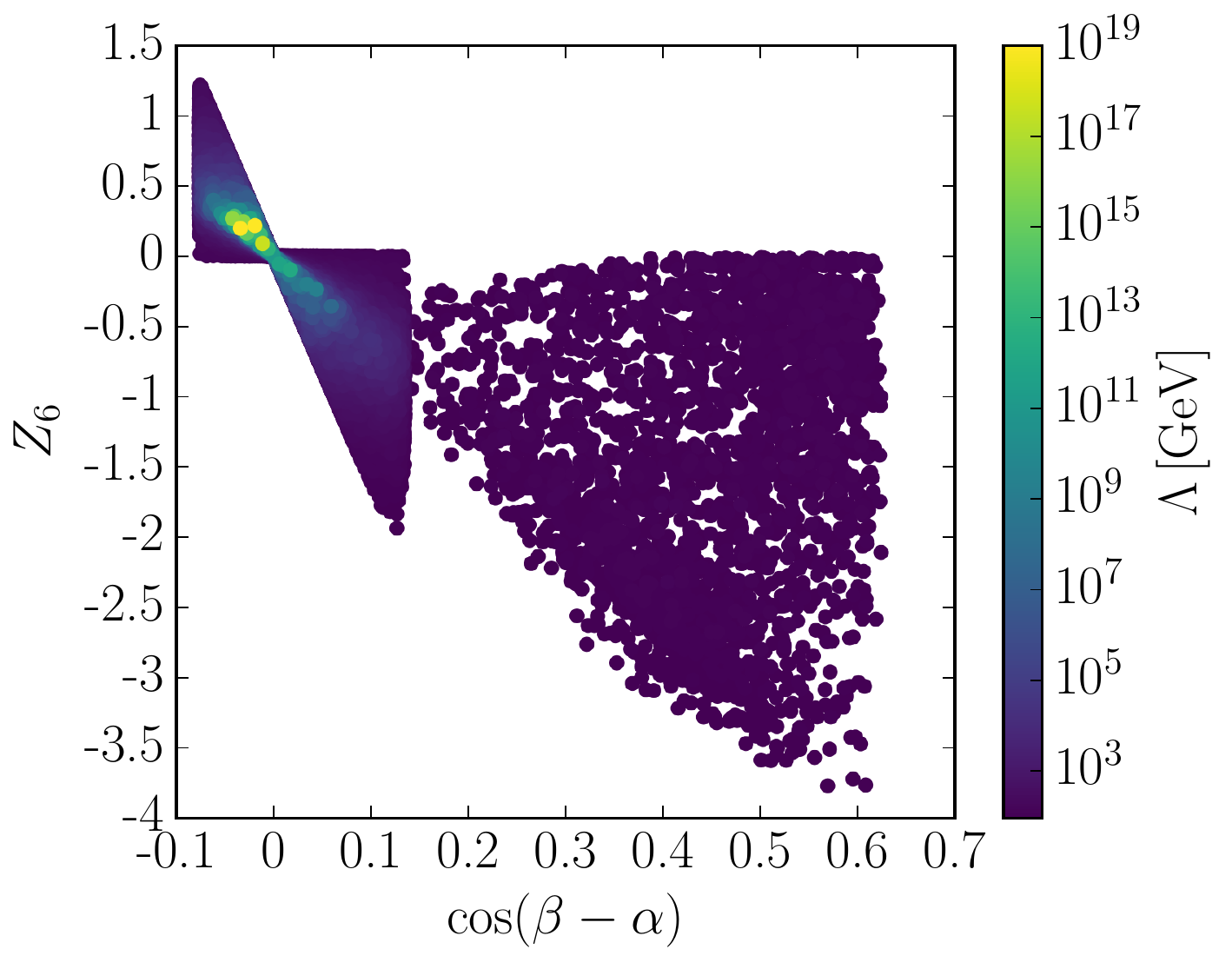}
  \caption{$Z_6$ as a function of $\cos (\beta - \alpha)$ in type I (left) and type II (right) colour coded with
  the cut-off scale.
    }\label{fig:Z6Z1_I}
\end{figure}

This leads us to a discussion on the alignment limit, given the plots
shown in Fig.~\ref{fig:Z6Z1_I}.
There we plot the values of the $Z_6$ Higgs-basis coupling (defined in
Eqs.~(\ref{higgspotHB}) and (\ref{eq:Z6})),
for both type I and type II models, vs. $\cos (\beta - \alpha)$. The
``cloud" of points in the type II plot, with large values of $\cos
(\beta - \alpha)$, corresponds to the {\em wrong sign limit} in that
model, to be discussed shortly. As can be seen in the left plot of
Fig.~\ref{fig:Z6Z1_I}, values of $Z_6$ very close to zero are
possible, particularly with the model valid up to
the Planck scale. However,
even with $Z_6$ of very low magnitude, $\cos (\beta - \alpha)$ clearly
still varies in a very wide range, and therefore we are {\em not} in the alignment limit.
Notice the striking difference with the same plot for model type II --
there, alignment indeed  corresponds to small $Z_6$. The difference in
behaviour between the two models is clearly due to the lower bound on
the charged Higgs mass. Therefore, very small values of the quartic couplings may not
be enough to make us reach the alignment limit -- although for the type II
model small $Z_6$ is indeed sufficient for alignment, the same cannot
be said for type I. For this model, small $Z_6$ needs to be
complemented by a large enough bound on the charged Higgs mass so that
one reaches alignment. We therefore would argue that the alignment
limit condition $|Z_6| \ll 1$ is a {\em necessary} condition, albeit
not a {\em sufficient} one. Still, there are certainly values of $Z_6$
closer to zero for which alignment would occur independently of the
values of the scalar masses, because the matrix in eq.~\eqref{eq:mhhb}
becomes very nearly diagonal for increasingly smaller $Z_6$.

\begin{figure}[t!]
  \centering
  \includegraphics[width=0.47\linewidth]{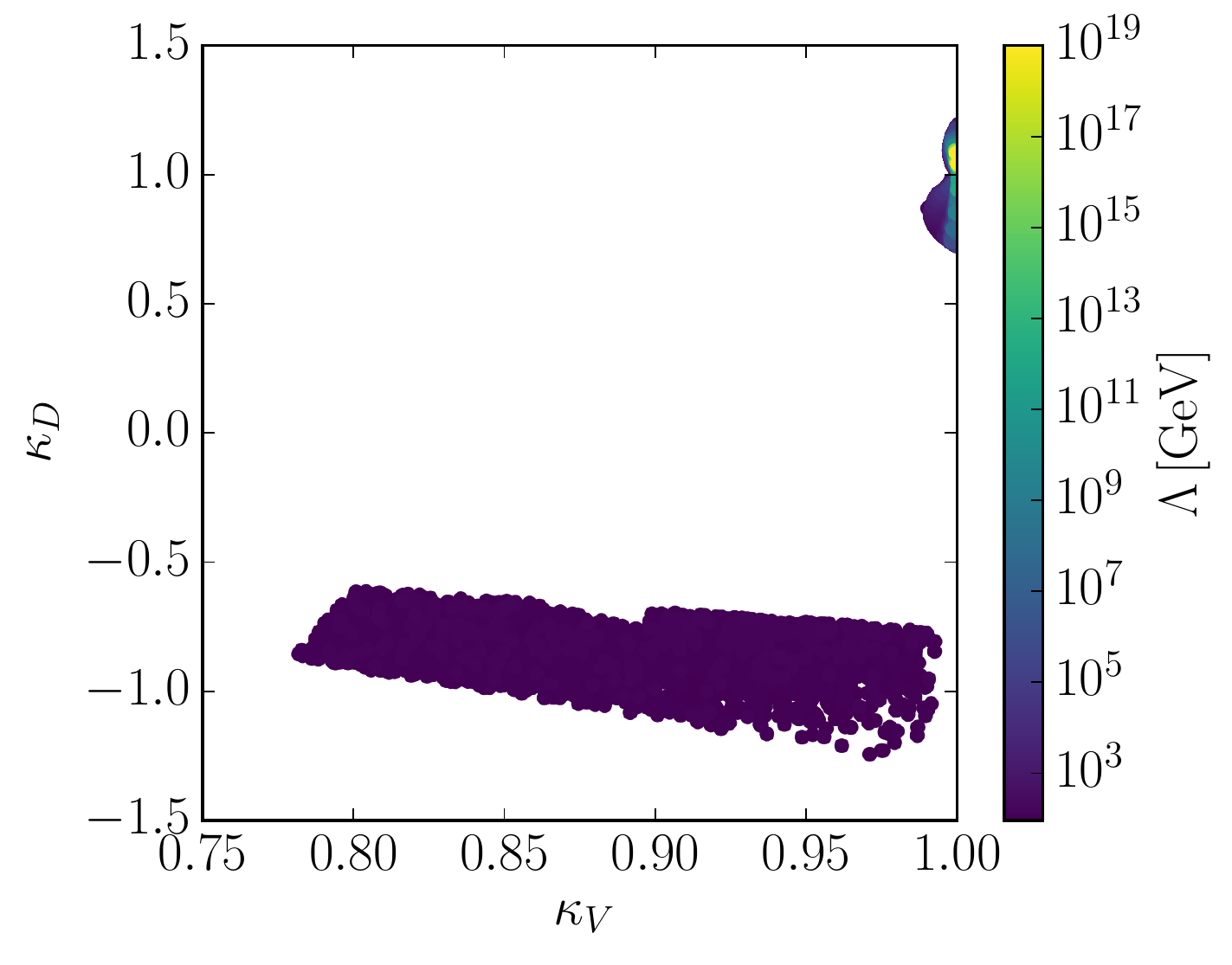}
  \includegraphics[width=0.37\linewidth]{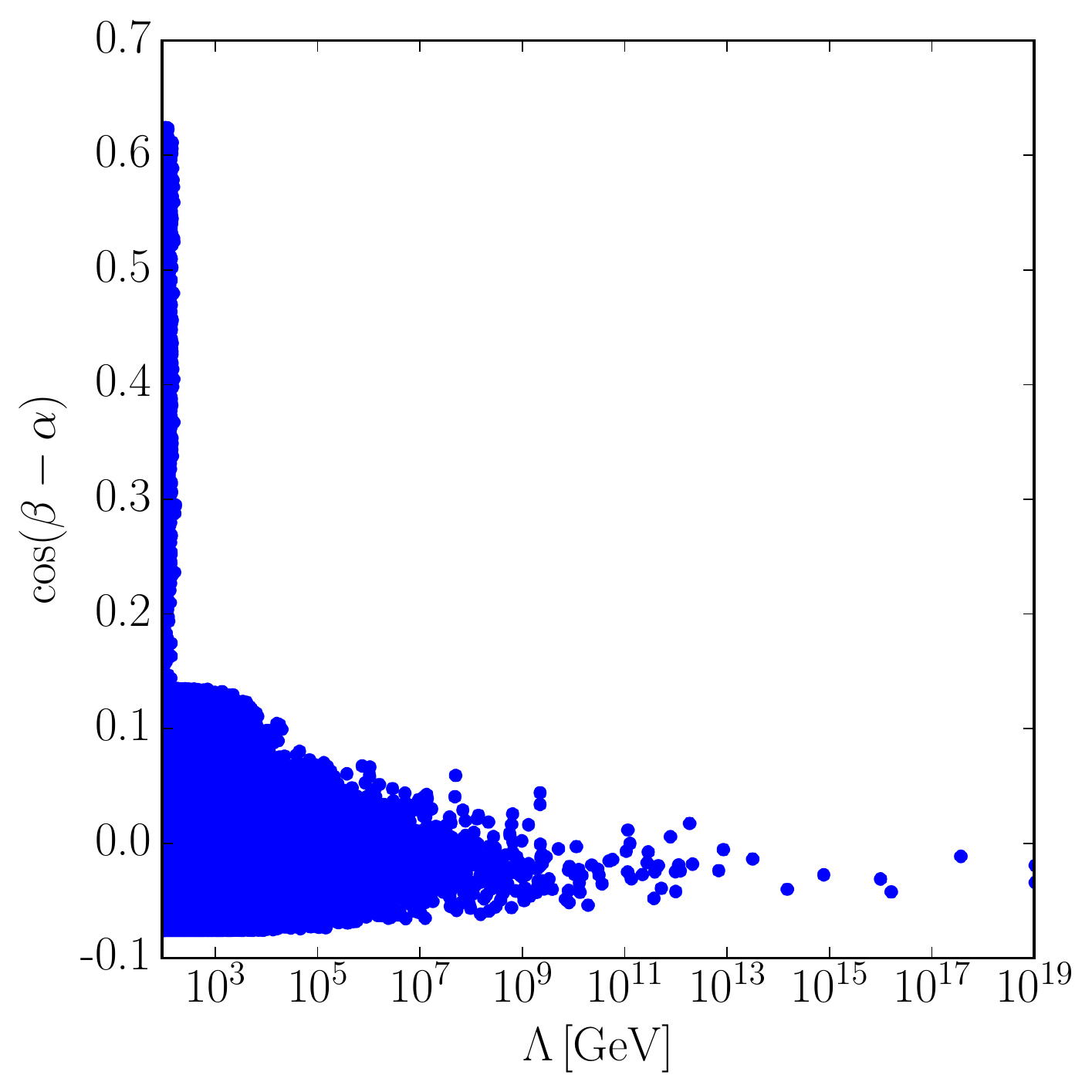}
  \caption{Left: $\kappa_D$ as a function of $\kappa_V$ in the type II 2HDM. The colour bar indicates
  the cut-off scale at which the model is no longer valid. Right:
  $\cos (\beta - \alpha)$ as a function of the cut-off scale
  $\Lambda$.
  }
\label{fig:wrong}
\end{figure}
In previous works~\cite{Ferreira:2014naa,Ferreira:2014dya} two of the
authors have discussed the wrong-sign limit of the 2HDM.
We define
\begin{equation}
\kappa_i^2=\frac{\Gamma^{\scriptscriptstyle {\rm 2HDM}}  (h \to i)}{\Gamma^{\scriptscriptstyle {\rm SM}} (h \to i)}
\end{equation}
which at tree-level is just the ratio of the couplings $\kappa_i=g_{i}^{\scriptscriptstyle {\rm 2HDM}}
  /g_{i}^{\scriptscriptstyle {\rm SM}} $ and for the  $h W^+ W^-$ coupling  reads
\begin{equation}
\kappa_{W}^2= \frac{\Gamma^{\scriptscriptstyle {\rm 2HDM}}  (h \to W^+ W^-)}{\Gamma^{\scriptscriptstyle {\rm SM}}  (h \to W^+ W^-)}=
\left( \frac{g_{\scriptscriptstyle  h W^+ W^-}^{\scriptscriptstyle
      {\rm 2HDM}} }{g_{ \scriptscriptstyle h W^+
      W^-}^{\scriptscriptstyle {\rm SM}}} \right)^2=\sin^2 (\beta -
\alpha) \;.
\end{equation}
Representing the down-type (up-type)
fermion final states by $\kappa_D$ ($\kappa_U$), the wrong-sign limit
is defined by $\kappa_D \kappa_V < 0$, that is, the down-type
couplings have a minus relative sign to the SM couplings. Other
wrong-sign limits could be defined but they are all excluded by
experiment~\cite{Ferreira:2014naa,Ferreira:2014dya}.
For completeness, the wrong sign limit is only possible in the type II
model, and for the light Higgs scenario it implies $\alpha > 0$, which
leads to sizeable values of $\cos (\beta - \alpha)$.

On the left panel of Fig.~\ref{fig:wrong} we present a plot of $\kappa_D$ as a function
of $\kappa_V$, where all the theoretical and experimental constraints
have been imposed, and the colour code indicates the scale up to which
the model is valid. Notice that the only region for which the model is
valid to higher scales corresponds to $\kappa_D > 0$ -- thus the validity
of the 2HDM up to high scales eliminates the wrong sign limit. To
enforce this conclusion, consider the right plot in
Fig.~\ref{fig:wrong}, wherein we show the values of $\cos (\beta
-\alpha)$ as a function of the cut-off scale $\Lambda$. We clearly see that above about
half a TeV the theory is valid only for points with very low values of $\cos (\beta -\alpha)$. 
Since the wrong sign limit can only occur with
sizeable values of $\cos (\beta
-\alpha)$~\cite{Ferreira:2014naa,Ferreira:2014dya}, it is therefore
excluded if one requires the model to be valid up to scales as low as
1 TeV.

In fact, it is easy to understand why the wrong sign limit is excluded by the high scale behaviour of
type II, if one remembers the results from Fig.~\ref{fig:massdiff_II_no_bounds}. Our analysis of those
plots led us to conclude that validity up to high scales of type II
placed us definitely in the decoupling
regime -- and as was shown in~\cite{Ferreira:2014naa}, the wrong sign limit corresponds to a {\em
non-decoupling} regime, wherein the charged Higgs boson has an irreducible contribution to observables
such as the diphoton width of the SM-like Higgs boson or the gluon fusion cross section.

\section{The heavy Higgs scenario \label{sec:res2}}

In this section we will discuss the scenario where the heavier of the
two CP-even Higgs bosons is the discovered 125 GeV scalar~\cite{Ferreira:2012my}. In this scenario
decoupling  cannot happen because the lightest scalar mass is constrained to be below 125 GeV. Hence, one
would expect that the theory would only be valid up to a certain scale, at least for the type II
model -- as we have seen, validity of type II to very high scales is only possible if the model is
in the alignment limit and all extra scalar masses are heavy. Recall also that what makes possible
the validity of the model up to the Planck scale is not only the fact that there is a new scalar, relative
to the SM, but also that the mass scale is driven by the $M$ parameter
and not by the quartic couplings. In what follows, all relevant experimental bounds were taken into
 account when generating the data samples. In particular, since we are considering the possibility of
 scalars lighter than 125 GeV, the LEP constraints~\cite{Barate:2003sz,Schael:2006cr} assume a
 special relevance in what follows.

\begin{figure}[h!]
  \centering
  \includegraphics[width=0.49\linewidth]{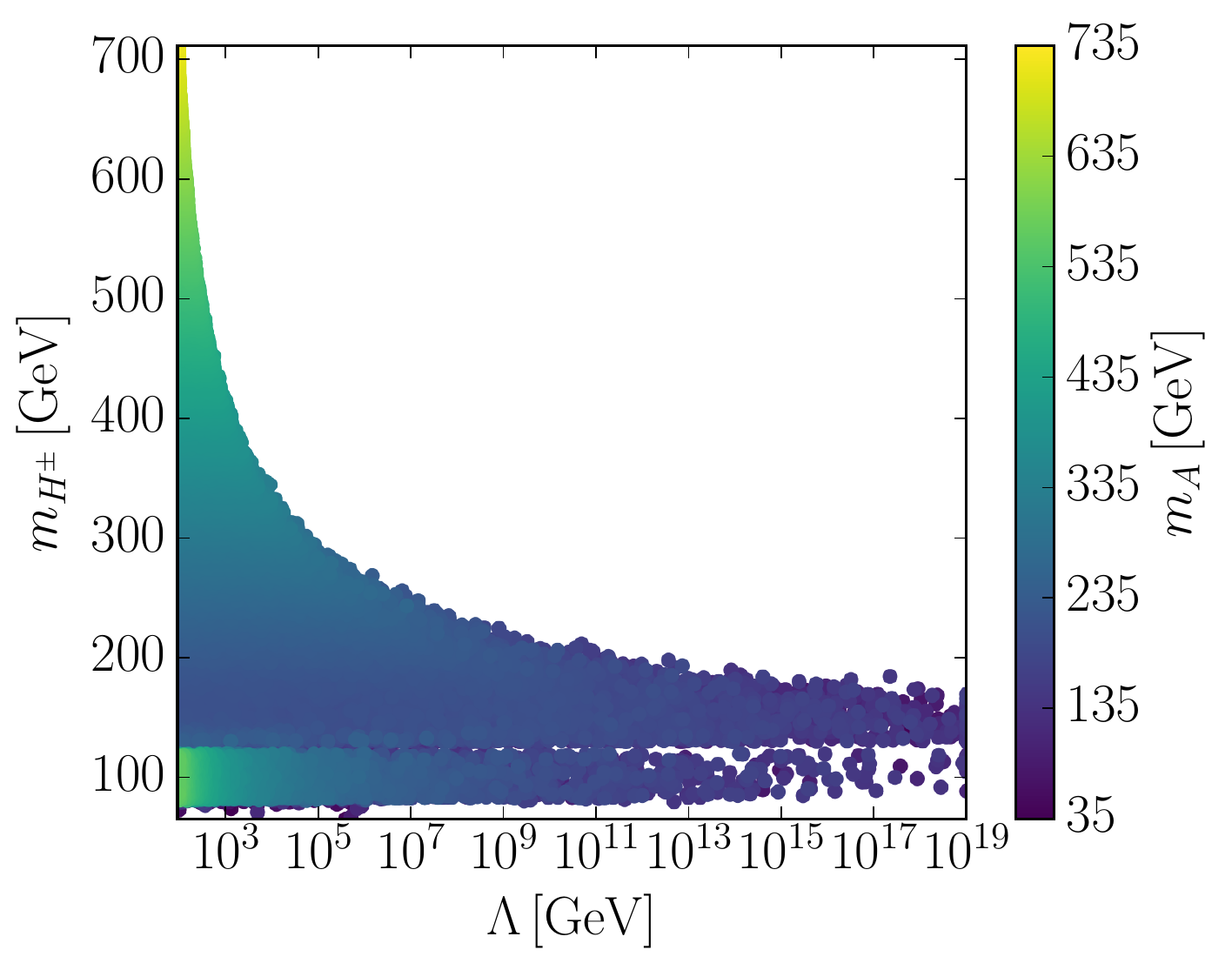}
  \includegraphics[width=0.38\linewidth]{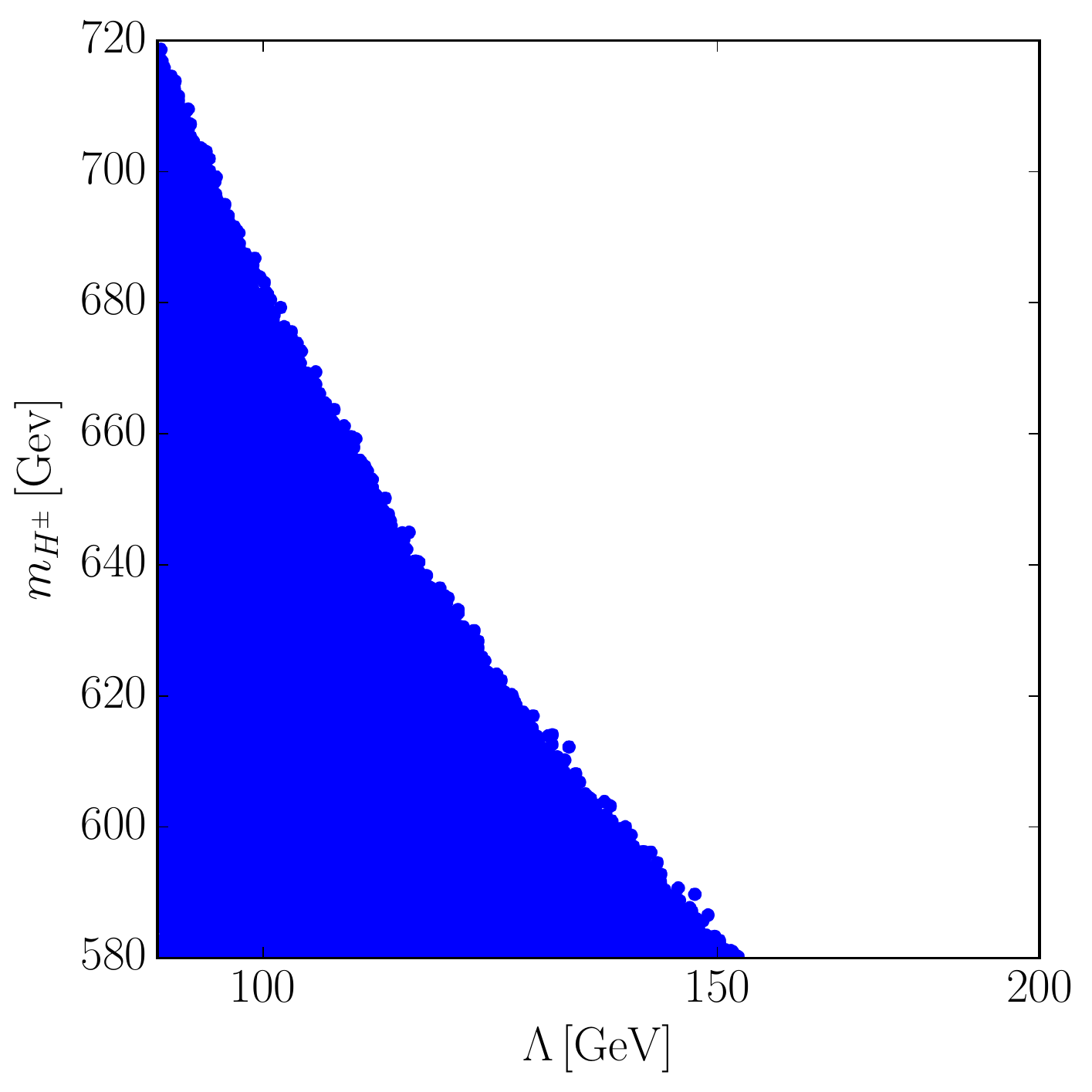}
  \caption{Charged Higgs mass as a function of the cut-off scale
    $\Lambda$, for type I (left) and type II (right). The colour coded
    bar in the left plot shows the value of
    $m_A$.
    }\label{fig:heavychargedscale}
\end{figure}
In Fig.~\ref{fig:heavychargedscale} we present the charged Higgs mass as a function of the cut-off scale
for type I (left) and type II (right). In the type I model it can be clearly seen that
there is a range of masses, both for the charged and pseudoscalar, that survive up to the Planck scale.
In particular, acceptable charged masses are above the LEP bound but
below 200 GeV.
Therefore the type I model survives up to the Planck scale if the charged Higgs boson and/or
the pseudoscalar are light. On the other hand, in type II, the stringent bound of 580 GeV not only
precludes the possibility of a light charged scalar but the theory ceases to be valid already
at $\Lambda \approx 150$~GeV. Even if we consider only the appearance
of Landau poles the type II model is valid only up to about 1.5 TeV.

\begin{figure}[t!]
  \centering
  \includegraphics[width=0.47\linewidth]{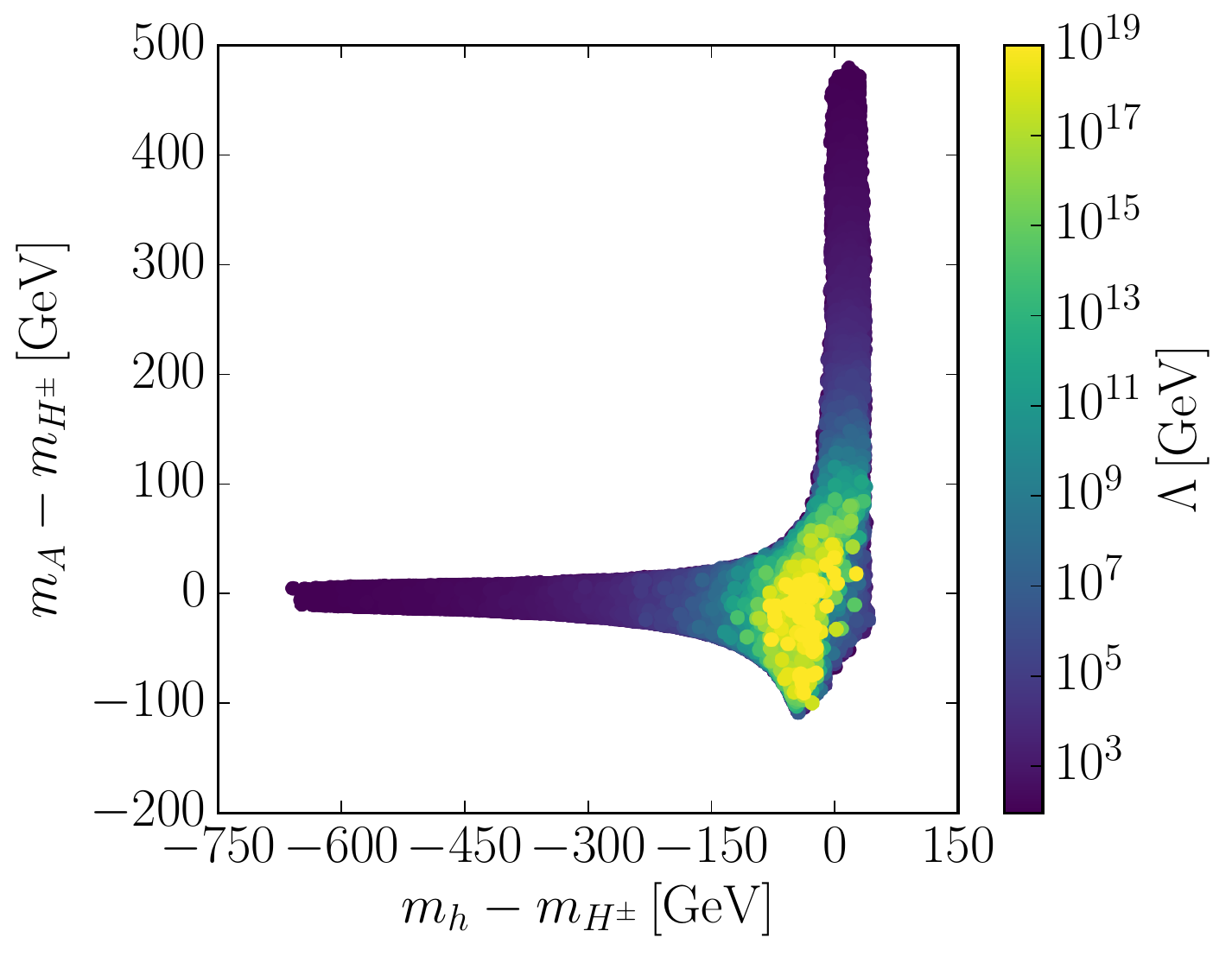}
  \includegraphics[width=0.47\linewidth]{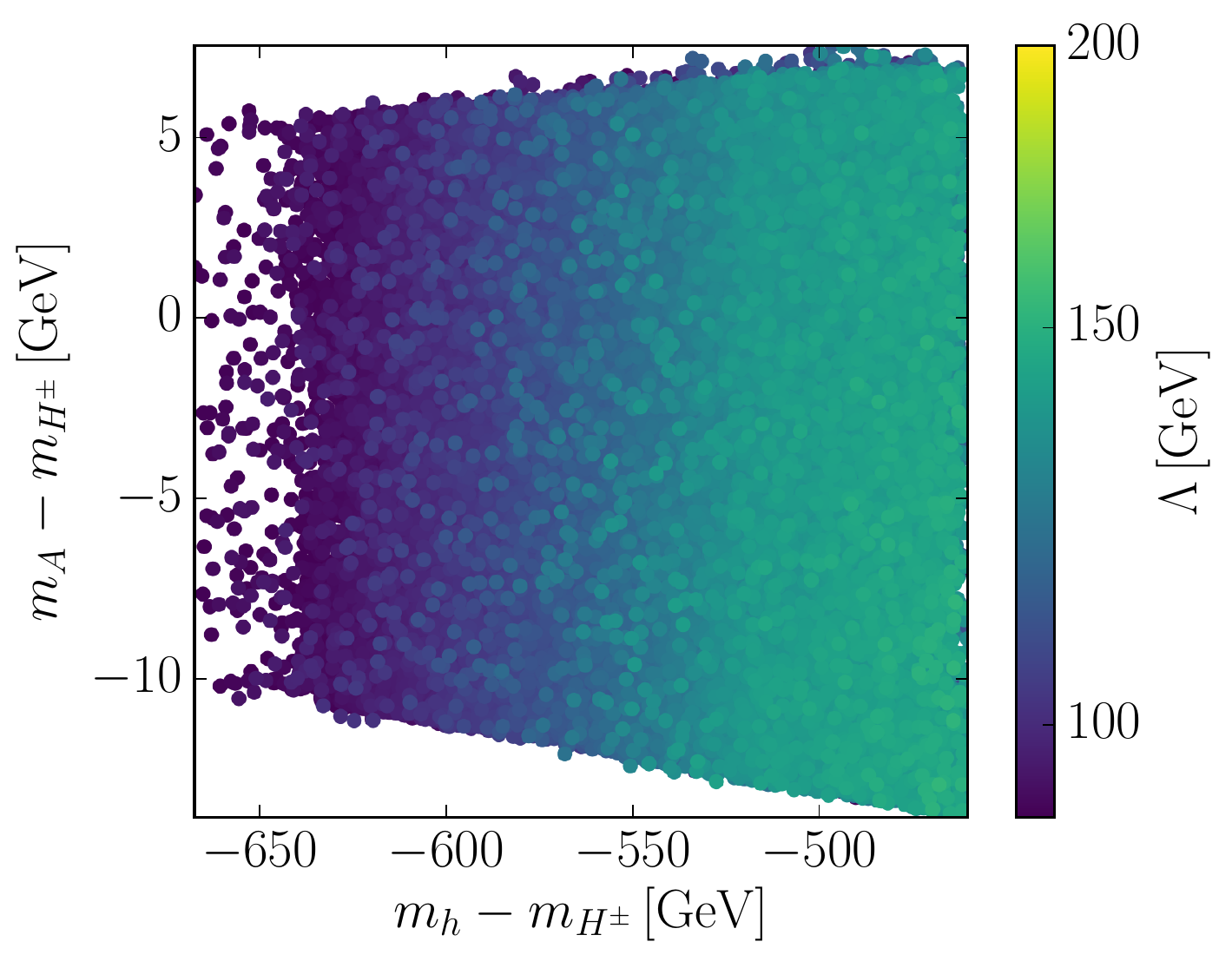}
  \caption{The mass difference $m_A - m_{H^\pm}$ vs. $m_h - m_{H^\pm}$ for type I (left) and type II (right). The colour code shows the cut-off
  scale $\Lambda$.
    }\label{fig:heavymassdiffscale}
\end{figure}
In fact, in type I, it is not only the charged Higgs boson that needs to be light for the model to
be valid up to the Planck scale. In Fig.~\ref{fig:heavymassdiffscale} we show $m_A - m_{H^\pm}$
vs. $m_h - m_{H^\pm}$ for type I (left) and type II (right). The
colour code shows at which energy either a Landau pole occurs or one
of the theoretical conditions is violated.
It is clear that for type I to be valid up to the Planck scale the pseudoscalar also needs to be light,
 and in fact all mass differences have to be below 100 GeV. We have also checked that the value of $M$ has to be
 of the same order and lies between 40 GeV and 120 GeV.
The cut-off scale has no major influence on the range\footnote{Note that in the
heavy Higgs scenario the alignment limit is attained for $\sin (\beta
-\alpha) \approx 0$.} of $\sin (\beta -\alpha)$ nor on the range
of $\tan \beta$. In fact, there is only a slight reduction in the allowed region with a slight increase in
the lower bound of $\tan \beta$, which moves closer to $\tan \beta > 4$.

The situation is radically different in the type II model, as can be appreciated from
the plot on the right in Fig.~\ref{fig:heavymassdiffscale}. We may conclude from that plot that,
due to the bound on the charged Higgs mass, the type II model barely survives up to a scale of 200 GeV.
Once again, this is due to the fact that validity of the type II model up to high scales eliminates
the possibility of non-decoupling regimes, of which the heavy Higgs scenario is certainly one. However,
there are regions in the parameter space of type I, in the heavy Higgs scenario, valid up to the Planck scale,
which is certainly surprising.
The conditions for this to happen are
$m_{H^\pm} \approx m_A \approx M$ and all masses below about $200$ GeV. There is no preferred value of the
lightest scalar mass even when one requires the model to be valid up
to the Planck scale.\footnote{It is easier  to fit the existing data with $m_h$ above roughly 62.5 GeV, but that is due to possible $H\rightarrow hh$ decays
potentially enlarging the branching ratio of $H$
beyond what is acceptable to be compatible with the
  LHC data on the Higgs rates.} And as in the case of
the light Higgs scenario, the running does not force the type I model
to move further close to the alignment limit.

\section{High scale behaviour and 2HDM symmetries}
\label{sec:res3}

The potential presented in Eq.~(\ref{higgspot}) is $\mathbb{Z}_2$
symmetric, softly broken by the $m_{12}^2$ (or $M$) term. The potential has an exact $\mathbb{Z}_2$ symmetry when $M=0$. We have shown in
Fig.~\ref{fig:explainmassdiff_no_bounds} (left) $M$ as a function of
the cut-off scale $\Lambda$ in the type II 2HDM while in Fig.~\ref{fig:symZ2} we show
the same plot but now for type I. These plots allow us to analyse the possibility of the
$\mathbb{Z}_2$ symmetry actually being an {\em exact} symmetry, unbroken even softly. To do that,
we simply need to investigate the possibility of $M$ - and therefore $m_{12}^2$ - being equal to zero
while the 2HDM is still valid up to high energy scales.

Now, it is quite clear from Fig.~\ref{fig:explainmassdiff_no_bounds}
(left) that $M = 0$ is only
a possibility for type II if the validity scale of the model is well
below 1 TeV\footnote{We differ
from previous calculations~\cite{Chakrabarty:2014aya,
  Das:2015mwa,Chowdhury:2015yja} due
to the higher charged Higgs mass bound, which had used
a value of 350 GeV.} -- thus one can
conclude that a type II model with an exact $\mathbb{Z}_2$ symmetry is
already strongly ruled out.
For the type I model, Fig.~\ref{fig:symZ2} shows that $M = 0$ is a
possibility for theories valid up
to scales of roughly 10 TeV, but no more than that. Thus an exact
$\mathbb{Z}_2$ symmetry is a possibility
for a type I model, but only if new physics is present at $\sim$ 10
TeV. Once again, the explanation for this
is due to the fact that most non-decoupling regimes are excluded if one
considers the 2HDM valid to very
high energies --- and the motivation for introducing a soft breaking
term in the potential
is indeed to allow for the possibility of a decoupling regime
occurring. Still, as we have showed
for type I, although there is no decoupling in the heavy Higgs
scenario, it is only the existence of the
term $m_{12}^2$ that allows the model to survive up to the Planck scale.
With $m_{12}^2$ absent,
the 2HDM becomes a non-decoupled theory, where some quantities (such as the diphoton width)
never conform to SM
expectations~\cite{Kanemura:2004mg}.

We were also able to investigate the possibility of the 2HDM possessing a $U(1)$
Peccei-Quinn symmetry~\cite{Peccei:1977hh}. This symmetry would imply $\lambda_5 = 0$
and in its exact form, also $m_{12}^2 = 0$. Again one can break it with a soft breaking
term and the symmetry is extended to the Yukawa sector in the same manner as the
$\mathbb{Z}_2$ model. But in this model the pseudoscalar $A$ is massless if no
soft breaking term $m_{12}^2$ is introduced, so we are not interested in the
exact symmetry scenario. And since $\lambda_5 = 0$ is enforced by a symmetry, it is
a fixed point in the RGE running of the quartic couplings. Thus we may ask if,
given all collider constraints existent, and requiring the model to be valid up to
very high scales, the RGE running is making the $\mathbb{Z}_2$ model tending to the
Peccei-Quinn one. This would happen if only values of $\lambda_5$  close to
zero -- more generically, of magnitude much smaller than the remaining couplings --
would survive the running. However, the results show that all $\lambda_i$ have similar
allowed ranges at the Planck scale. This was checked for all scenarios presented in this work.
We have also verified, with a separate data sample generated specially for this verification,
that for a type I Peccei-Quinn model satisfying all LHC constraints, there are regions of
parameter space for which the theory is valid all the way up to the Planck scale. In fact,
the results for the $U(1)$ model are, in this regard, indistinguishable from those of
the $\mathbb{Z}_2$ case.

%%%%%%%%%%%%%%%%%%%%%%%%%%%%%%%%%%%%%%%%%%%%%%%%%%%%%%%
\section{Conclusions \label{sec:concl}}

We have analysed the high scale behaviour of a softly broken $\mathbb{Z}_2$ symmetric 2HDM
focusing on two particular Yukawa types, type I and type II. If the lightest CP-even
scalar is the 125 GeV one, there are regions of the parameter space for both types
that survive up to the Planck scale. This is a confirmation of many previous studies
in the literature. There are, however, new and quite interesting results, some of them
unexpected that we will now discuss.

One of the most interesting conclusions of our study is that for the model to be close to the alignment
limit it is enough to require it to be valid up to about 1 TeV and at the same time
to have one of the scalar masses above about 500 GeV (in the specific
case of type II $B$-physics bounds force the charged Higgs mass to be above 580 GeV).
No other bounds need to be considered to reach this limit. As the scale up to which we want
the model to be valid increases, the allowed region of parameter space moves closer and closer to alignment.
Therefore, alignment is reached via decoupling -- at least a large (above roughly 500 GeV)
scalar mass is required.

On the contrary, we have shown that for type I, for which there are no strong bounds on the scalar masses,
validity up to the Planck scale will not imply alignment, if the masses are low enough. In fact, even when all experimental constraints are considered, the type I model can be far from alignment, except for large scalar
masses where we then recover the results obtained for type II. In this sense, validity up to higher scales
(as low as 1 TeV in certain cases), complemented with one sufficiently large scalar mass (above 500 GeV)
implies alignment in the 2HDM, a phenomenon we might call ``radiative alignment". In this sense,
alignment in the 2HDM is therefore fundamentally caused by the behaviour of the theory at high scales,
instead as, for instance, the occurrence of symmetries -- of which the inert doublet
model~\cite{Ma:1978,Barbieri:2006,LopezHonorez:2006,Krawczyk:2013jhep} is a prime example; another
possibility would be the model developed in
Ref.~\cite{Draper:2016cag}.

The validity up to high scales of the scenario where the heaviest CP-even scalar is the 125 GeV Higgs
was analysed here for the first time. Interestingly, we have shown that there are regions of the parameter
space where a type I model, in the heavy Higgs scenario, is valid up to the Planck scale.
That is, a model with no decoupling limit can be valid up to the Planck scale.
The most interesting point to note is that also in this case it is the soft parameter
$M$ that sets the mass scale for validity at
high energies. In fact, all masses have to be below about 200 GeV to ensure that
the model does not require new physics up to the Planck scale. On the contrary, and
again due to the bound on the charged Higgs mass, the type II model in the heavy
Higgs scenario does not survive, even to a scale of just a few hundred GeV.
In all these scenarios we should highlight the important role played by the parameter $M$.
Indeed, because the quartic couplings become increasingly small, all models that survive up
to the Planck scale need a non-zero $M$ and all masses are of the order of $M$ (except for the 125
GeV Higgs boson).

The previous result is even more interesting when combined with knowledge
that non-decoupling scenarios in type II, such as the wrong sign limit, will not survive
to scales as low as a few TeV. As the quartic couplings increase with energy, they have to be quite small
to survive the running. Hence, any non-decoupling regime that needs large quartic couplings
will not survive to high scales.

Finally we have shown that the model does not approach the exact $\mathbb{Z}_2$ symmetry nor the softly
broken $U(1)$ symmetry when the validity of the theory is required up to the Planck scale. In fact,
the value of $M^2 = 0$ is disallowed for type II already at a scale well below 1 TeV while
for type I it happens at a scale of about 10 TeV. As for the softly broken $U(1)$ where
$\lambda_5 =0$, we have shown that requiring the validity of the model up to the Planck
scale forces all the quartic couplings to be small so that $\lambda_5$ behaves just like
the other quartic couplings.

%%%%%%%%%%%%%%%%%%%%%%%%%%%%%%%%%%%%%%%%%%%%%%%%%%%%%%%
%%%%%%%%%%%%%%%%%%%%%%%%%%%%%%%%%%%%%%%%%%%%%%%%%%%%%%%
\section*{Appendix}
\setcounter{equation}{0}
\begin{appendix}

%%%%%%%%%%%%%%%%%%%%%%%%%%%%%%%%%%%%%%%%%%%%%%%%%%%%%%%
\section{RGEs for the 2HDM \label{RGE}}
\label{app:rge}

The one-loop RGEs for the gauge couplings, Yukawa couplings and $\lambda$'s are taken from \cite{Branco:2011iw}.
The one-loop RGEs for the quadratic parameters $m_{ij}^ 2$ from eq.~\eqref{higgspot} are taken
from \cite{Haber:1993an}. We define
\begin{align}
	\beta_{x} &= 16\pi^2 \frac{\partial x}{\partial \ln \mu} \;.
\end{align}
The RGEs for the
$U(1)_Y$, $SU(2)_L$ and $SU(3)$
gauge couplings $g$, $g'$ and $g_s$, respectively, are given as
\begin{align}
	\beta_{g_s} &= -7 g_s^3 \,,\\
	\beta_{g} &= -3 g^3 \,,\\
	\beta_{g'} &= 7g'^3 \,.\\
\end{align}
For the Yukawa sector (see below Eqs.~(\ref{eq:Yu})--~(\ref{eq:Ye}) for the definition of the
Yukawa matrices $Y_x$) we have in type I
\begin{align}
	\beta_{Y_u} &= a_u Y_u + T_{22} Y_u - \frac{3}{2} \left( Y_d Y_d^\dagger - Y_u Y_u^\dagger\right) Y_u \,,\\
	\beta_{Y_d} &= a_d Y_d + T_{22} Y_d + \frac{3}{2} \left(Y_d Y_d^\dagger - Y_u Y_u^\dagger\right)Y_d \,,\\
	\beta_{Y_e} &= a_e Y_e + T_{22} Y_e + \frac{3}{2} Y_eY_e^\dagger Y_e \,,
\end{align}
and in type II
\begin{align}
	\beta_{Y_u} &= a_u Y_u + T_{22} Y_u + \frac{1}{2} \left(Y_dY_d^\dagger + 3 Y_uY_u^\dagger\right)Y_u \,,\\
	\beta_{Y_d} &= a_d Y_d + T_{11} Y_d + \frac{1}{2} \left(Y_uY_u^\dagger + 3 Y_d Y_d^\dagger\right)Y_d \,,\\
	\beta_{Y_e} &= a_e Y_e + T_{11} Y_e + \frac{3}{2} Y_eY_e^\dagger Y_e \,,
\end{align}
with
\begin{align}
	a_d & =-8g_s^2 - \frac{9}{4} g^2 -\frac{5}{12} g'^2 \,,\\
	a_u &= -8g_s^2 - \frac{9}{4} g^2 - \frac{17}{12} g'^2 \,,\\
	a_e &= - \frac{9}{4} g^2 - \frac{15}{4} g'^2 \,.
\end{align}
For type I we define
\begin{align}
	T_{11} &= 0 \,,\\
	T_{22} &= 3 Y_u^\dagger Y_u + 3 Y_d^\dagger Y_d + Y_e^\dagger Y_e \,,
\end{align}
and for type II we have
\begin{align}
	T_{11} &= 3 Y_d^\dagger Y_d + Y_e^\dagger Y_e \,,\\
	T_{22} &= 3 Y_u^\dagger Y_u \,.
\end{align}
For the quartic couplings we have in type I
\begin{align}
	\beta_{\lambda_1} &= 12\lambda_1^2 +4\lambda_3^2 + 4 \lambda_3\lambda_4 + 2\lambda_4^2 + 2 \lambda_5^2 + \frac{9}{4} g^4+ \frac{3}{2} g^2g'^2  + \frac{3}{4} g'^4 - 4\gamma_1 \lambda_1 \,,\\
	\beta_{\lambda_2} &= 12\lambda_2^2 +4\lambda_3^2 + 4\lambda_3\lambda_4 + 2\lambda_4^2 +2\lambda_5^2 + \frac{9}{4} g^4+ \frac{3}{2} g^2g'^2  + \frac{3}{4} g'^4 - 4\gamma_2 \lambda_2 \notag \\ &\quad -12\mathrm{Tr}\left[Y_d^\dagger Y_d Y_d^\dagger Y_d + Y_u^\dagger Y_u Y_u^\dagger Y_u\right] - 4 \mathrm{Tr}\left[Y_e^\dagger Y_e Y_e^ \dagger Y_e\right] \,,\\
	\beta_{\lambda_3} &= \left(\lambda_1+\lambda_2\right)\left(6\lambda_3+2\lambda_4\right)+4\lambda_3^2 + 2\lambda_4^2 + 2 \lambda_5^2 + \frac{9}{4} g^4 - \frac{3}{2} g^2g'^2 + \frac{3}{4} g'^4 - 2\left( \gamma_1 + \gamma_2 \right)\lambda_3\,,\\
	\beta_{\lambda_4} &= 2\left(\lambda_1+\lambda_2\right)\lambda_4  + 8\lambda_3\lambda_4 + 4 \lambda_4^2 + 8 \lambda_5^2  - 2\left(\gamma_1+\gamma_2\right)\lambda_4 + 3g^2 g'^2\,,\\
	\beta_{\lambda_5} &= 2\left( \lambda_1 + \lambda_2 + 4
                            \lambda_3 + 6 \lambda_4 \right)\lambda_5 -
                            2\left(\gamma_1+\gamma_2\right) \lambda_5 \;,
\end{align}
and in type II
\begin{align}
	\beta_{\lambda_1} &= 12\lambda_1^2 +4\lambda_3^2 + 4 \lambda_3\lambda_4 + 2\lambda_4^2 + 2 \lambda_5^2 + \frac{9}{4} g^4+ \frac{3}{2} g^2g'^2  + \frac{3}{4} g'^4 - 4\gamma_1 \lambda_1 \notag \\ &\quad -12\mathrm{Tr}\left[Y_d^\dagger Y_d Y_d^\dagger Y_d \right] - 4 \mathrm{Tr}\left[Y_e^\dagger Y_e Y_e^\dagger Y_e\right] \,,\\
	\beta_{\lambda_2} &= 12\lambda_2^2 +4\lambda_3^2 + 4\lambda_3\lambda_4 + 2\lambda_4^2 +2\lambda_5^2 + \frac{9}{4} g^4+ \frac{3}{2} g^2g'^2  + \frac{3}{4} g'^4 - 4\gamma_2 \lambda_2 \notag \\ &\quad -12\mathrm{Tr}\left[Y_u^\dagger Y_u Y_u^\dagger Y_u\right]  \,,\\
	\beta_{\lambda_3} &= \left(\lambda_1+\lambda_2\right)\left(6\lambda_3+2\lambda_4\right)+4\lambda_3^2 + 2\lambda_4^2 + 2 \lambda_5^2 + \frac{9}{4} g^4 - \frac{3}{2} g^2g'^2 + \frac{3}{4} g'^4 - 2\left( \gamma_1 + \gamma_2 \right)\lambda_3 \notag \\ &\quad - 12 \mathrm{Tr}\left[Y_d^\dagger Y_d Y_u^\dagger Y_u\right] \,,\\
	\beta_{\lambda_4} &= 2\left(\lambda_1+\lambda_2\right)\lambda_4  + 8\lambda_3\lambda_4 + 4 \lambda_4^2 + 8 \lambda_5^2  - 2\left(\gamma_1+\gamma_2\right)\lambda_4 + 3g^2 g'^2  \notag \\ &\quad + 12 \mathrm{Tr}\left[Y_d^\dagger Y_d Y_u^\dagger Y_u \right]\,,\\
	\beta_{\lambda_5} &= 2\left( \lambda_1 + \lambda_2 + 4
                            \lambda_3 + 6 \lambda_4 \right)\lambda_5 -
                            2\left(\gamma_1+\gamma_2\right) \lambda_5 \;,
\end{align}
with
\begin{align}
	\gamma_1 &= \frac{9}{4} g^2 + \frac{3}{4} g'^2 - T_{11} \,,\\
 	\gamma_2 &= \frac{9}{4} g^2 + \frac{3}{4} g'^2 - T_{22} \,.
\end{align}
For the dimensionful couplings we have
\begin{align}
	\beta_{m_{11}^2} &= 6\lambda_1 m_{11}^2 + \left(4\lambda_3+2\lambda_4\right) m_{22}^2 - 2\gamma_1 m_{11}^2 \,,\\
	\beta_{m_{22}^2} &= \left(4\lambda_3 + 2\lambda_4\right)m_{11}^2 + 6 \lambda_2 m_{22}^2 - 2\gamma_2 m_{22}^2 \,,\\
	\beta_{m_{12}^2} &= \left(2\lambda_3+4\lambda_4 + 6\lambda_5\right) m_{12}^2 - \left(\gamma_1+\gamma_2\right)m_{12}^2 \,.
\end{align}
The RGEs for the VEVs are given by \cite{Sperling2013,Sperling2013a}
\begin{align}
	\beta_{v_1} &= \gamma_1 v_1 \,,\\
	\beta_{v_2} &= \gamma_2 v_2 \,.
\end{align}
Our starting values are given by
\begin{align}
	g_s &= \sqrt{4\pi \alpha_s} \,,\\
	g &= \frac{2m_W}{v} \,,\\
	g' &= 2 \frac{\sqrt{m_Z^2-m_W^2}}{v} \,,\\
	Y_u &= \frac{\sqrt{2}}{v_2} \begin{pmatrix}
		m_u &0&0\\0&m_c&0\\0&0&m_t
	\end{pmatrix} \,,\label{eq:Yu} \\
	Y_d &= \frac{\sqrt{2}}{v_d} V_{CKM} \begin{pmatrix}
	m_d &0&0\\0&m_s&0\\0&0&m_b
	\end{pmatrix} V_{CKM}^\dagger \,,\label{eq:Yd}\\
	Y_e &= \frac{\sqrt{2}}{v_e} \begin{pmatrix}
	m_e &0&0\\0&m_\mu&0\\0&0&m_{\tau}
	\end{pmatrix} \,, \label{eq:Ye} \\
	V_{CKM} &= 1_{3\times 3} \,,
\end{align}
where $\alpha_s = g_s^2/(4\pi)$ is the strong coupling
constant. In type I we have
\begin{align}
	v_e &= v_2 \,,\\
	v_d &= v_2 \,,\\
\end{align}
and in type II
\begin{align}
	v_e &= v_1 \,,\\
	v_d &= v_1 \,.
\end{align}
The fermion masses are chosen as \cite{Agashe:2014kda,Denner:2047636,LHCHXSWG,Dittmaier:2011ti}
\begin{align}
	m_u &= 0.1 \,\mathrm{GeV}\,,\\
	m_c &= 1.51 \,\mathrm{GeV}\,,\\
	m_t &= 172.5 \,\mathrm{GeV}\,,\\
	m_d &= 0.1 \,\mathrm{GeV}\,,\\
	m_s &= 0.1 \,\mathrm{GeV}\,,\\
	m_b &= 4.92 \,\mathrm{GeV}\,,\\
	m_e &= 0.510998928 10^{-3} \,\mathrm{GeV}\,,\\
	m_\mu &= 0.1056583715 \,\mathrm{GeV}\,,\\
	m_\tau &= 1.77682 \,\mathrm{GeV}\,.
\end{align}
The VEV is given by
\begin{align}
	G_F &= 1.1663787 \cdot 10^{-5} \,\mathrm{GeV}^{-2} \,,\\
	v &= \frac{1}{\sqrt{\sqrt{2} G_F }} \;,
\end{align}
and the strong coupling is
\begin{align}
	\alpha_s &= 0.119 \,.
\end{align}
The W and Z boson masses are given by \cite{Agashe:2014kda,Denner:2047636}
\begin{align}
	m_W &= 80.385 \,\mathrm{GeV} \,,\\
	m_Z &= 91.1876 \,\mathrm{GeV} \,.
\end{align}

\end{appendix}

%%%%%%%%%%%%%%%%%%%%%%%%%%%%%%%%%%%%%%%%%%%%%%%%%%%%%%%%%%%%

%%%%%%%%%%%%%%%%%%%%%%%%%%%%%%%%%%%%%%%%%%%%%%%%%%%%%%%
\vspace*{0.5cm}
\section*{Acknowledgments}
We would like to thank Jonas Wittbrodt for useful discussions
and for providing the 2HDM samples. PB acknowledges
financial support by the ``Karlsruhe School of Elementary Particle and
Astroparticle Physics: Science and Technology (KSETA)''.
MM acknowledges financial support from the DFG project “Precision Calculations in the Higgs Sector - Paving the Way to the New Physics Landscape” (ID: MU 3138/1-1).
PF and RS are supported in part by the National Science Centre, Poland, the
HARMONIA project under contract UMO-2015/18/M/ST2/00518.
\vspace*{0.5cm}
%%%%%%%%%%%%%%%%%%%%%%%%%%%%%%%%%%%%%%%%%%%%%%%%%%%%%%%

%%%%%%%%%%%%%%%%%%%%%%%%%%%%%%%%%%%%%%%%%%%%%%%%%%%%%%%%%%%%
\vspace*{1cm}
\bibliographystyle{h-physrev}
%\bibliography{run2hdm.bib}
\bibliography{run2hdm}
\end{document}